%% file: 00_main.tex
\pdfoutput=1
\documentclass{article}
\usepackage{arxiv}
\usepackage[utf8]{inputenc}
\usepackage[T1]{fontenc}
\usepackage{url}
\usepackage{booktabs}
\usepackage{babel}
\usepackage{amsfonts}
\usepackage{nicefrac}
\usepackage{microtype}
\usepackage{lipsum}
\usepackage{graphicx}
\usepackage{natbib}
\usepackage{doi}
\usepackage{abstract}
\usepackage{amsmath}
\usepackage{lineno}
\usepackage{authblk}
\usepackage{orcidlink}
\hypersetup{
    colorlinks,
    bookmarksopen,
    bookmarksnumbered,
    linkcolor=red,
    citecolor=red,
    urlcolor=red
}
\graphicspath{ {} }
\title{Spatiotemporal gender differences in urban vibrancy}
\author[1*]{Thomas Collins\,\orcidlink{0000-0003-1791-7002}\,}
\author[2, 3]{Riccardo Di Clemente\,\orcidlink{0000-0001-8005-6351}\,}
\author[4]{Mario Gutiérrez-Roig\,\orcidlink{0000-0002-8488-1954}\,}
\author[1, 2]{Federico Botta\,\orcidlink{0000-0002-5681-4535}\,}
\affil[1]{\footnotesize
Department of Computer Science\\
University of Exeter, Exeter\\
EX4 4QF, United Kingdom.}
\affil[2]{\footnotesize
Complex Connections Lab, Network Science Institute,\\
Northeastern University London\\
London, E1W 1LP, United Kingdom.}
\affil[3]{
The Alan Turing Institute\\
London, NW1 2DB, United Kingdom.}
\affil[4]{\footnotesize
Department of Mathematical Sciences\\
University of Essex, Colchester\\
CO4 3SQ, United Kingdom.}
\affil[*]{Corresponding author: trc207@exeter.ac.uk}
\date{}
\renewcommand{\headeright}{Preprint}
\renewcommand{\undertitle}{}
\renewcommand{\shorttitle}{Spatiotemporal gender differences in urban vibrancy}
\begin{document}
\maketitle
\input{01_abstract}
\input{02_introduction}
\input{03_data}
\input{04_methods}
\input{04_statistics}
\input{05_results}
\input{06_discussion}
\bibliographystyle{unsrtnat}
\bibliography{00_main}
\appendix
\renewcommand{\thesection}{Supplementary Information \arabic{section}}
\renewcommand{\thefigure}{SI~\arabic{figure}}
\setcounter{figure}{0}
\renewcommand{\thetable}{SI~\arabic{table}}
\setcounter{table}{0}
\input{07_supp}
\end{document}

%% file: 01_abstract.tex
\renewcommand{\abstractname}{}
\begin{abstract}
Urban vibrancy is the dynamic activity of humans in urban locations. It can vary with urban features and the opportunities for human interactions, but it might also differ according to the underlying social conditions of city inhabitants across and within social surroundings. Such heterogeneity in how different demographic groups may experience cities has the potential to cause gender segregation because of differences in the preferences of inhabitants, their accessibility and opportunities, and large-scale mobility behaviours. However, traditional studies have failed to capture fully a high-frequency understanding of how urban vibrancy is linked to urban features, how this might differ for different genders, and how this might affect segregation in cities. Our results show that (1) there are differences between males and females in terms of urban vibrancy, (2) the differences relate to `Points of Interest` as well as transportation networks, and (3) there are both positive and negative `spatial spillovers` existing across each city. To do this, we use a quantitative approach using Call Detail Record data--taking advantage of the near-ubiquitous use of mobile phones--to gain high-frequency observations of spatial behaviours across Italy's seven most prominent cities. We use a spatial model comparison approach of the direct and `spillover` effects from urban features on male-female differences. Our results increase our understanding of inequality in cities and how we can make future cities fairer.\par
\end{abstract}

%% file: 02_introduction.tex
\section{Introduction}\label{sec:introduction}
As the world continues to urbanize at an unprecedented rate, the lives of city inhabitants are transforming, with both unprecedented opportunities but also growing challenges and complexities that cannot be ignored. The United Nations reported that, by 2050, $68 \%$ of the world's population will be living in cities~\citep{undesaWorldUrbanizationProspects2019} and that, while urban populations are increasing, rural populations are in decline. This trend toward urban life is thought to be related to economic development alongside changes in social organisation~\citep{harrisUrbanisationEconomicDevelopment1990a, bettencourtGrowthInnovationScaling2007a}, how humans use land~\citep{liuGeographyHumanActivity2020}, and the drastic changes in the patterns of collective human behaviour~\citep{thomasUrbanisationDriverChange2008, bettencourtGrowthInnovationScaling2007a}. Rapid urban growth is thought to make cities more innovative and generate wealth but can cause large-scale social issues for people and communities. These include reduced housing affordability~\citep{nelsonLinkGrowthManagement2002}, environmental degradation~\citep{elarabyUrbanGrowthEnvironmental2002}, high crime rates with negative effects on economics, education, and health~\citep{glaeserWhyThereMore1999a}, greater disease incidence~\citep{connollyExtendedUrbanisationSpatialities2021,Santana2022} and traffic congestion~\citep{zhengUrbanComputingConcepts2014a,bassolasHierarchicalOrganizationUrban2019a,Kalila2018,Xu2021}.\par
Tóth et al. (2021) reported that rapid city growth can increase segregation and inequality in urban areas~\citep{tothInequalityRisingWhere2021}. Indeed, in cities in the United States, life expectancy has generally increased in the middle classes whereas, in poorer classes, it has remained the same~\citep{borPopulationHealthEra2017}. Within the spatial structure of cities, some neighbourhoods have become differentially desirable. More expensive locations force lower-income inhabitants away, and in some cases, to the fringes of cities or areas with increased levels of criminality or poverty~\citep{wellerGentrificationDisplacementEffects2012}. This can generate a powerful reinforcement loop thought to block those wishing to move to the area and hinder social mobility further.\par
One way to understand and quantify socio-spatial segregation in cities has been to use traditional data, like a census. However, by being based only on where people live, such data only ever `scratch at the surface` regarding the quantification of the fascinating details of urban environments and the relationship to our social lives and our ‘quality-of-life‘~\citep{entwislePuttingPeoplePlace2007b, kingJaneJacobsNeed2013a}. Thus, city planners increasingly look to new technologies to study collective human behaviour and, especially, to characterise mobility patterns~\citep{steenbruggenDataMobilePhone2015}. Data on broad movement behaviours are now accessible due to widespread interaction with technological systems~\citep{gonzalezUnderstandingIndividualHuman2008} and computer technology can help to reveal patterns in human behaviour. The world's near-ubiquitous uptake of mobile phone technology and social media generates huge amounts of data on our behaviour and mobility ~\citep{lazerLifeNetworkComing2009a, vespignaniPredictingBehaviorTechnoSocial2009, salessesCollaborativeImageCity2013a, bottaQuantifyingCrowdSize2015b, seresinheQuantifyingLinkArt2016,  preisSensingGlobalTourism2020}. From shopping habits~\citep{diclementeSequencesPurchasesCredit2018a, bannisterRapidIndicatorsDeprivation2021a,Xu2019} to transportation~\citep{suRhythmStreetsStreet2022}, there is an unlimited array of uses afforded to us due to this new ability to track and record the movements of citizens. This new direction has provided an extraordinary new understanding of urban environments and cities~\citep{battyNewScienceCities2013a,panUrbanCharacteristicsAttributable2013a, bottaQuantifyingCrowdSize2015b,barthelemyStructureDynamicsCities2016a, bottaSearchArtRapid2020a}.\par
Mobile phone data can support the study of \emph{urban vibrancy} or \emph{urban vitality}, which measures the activity of urban environments~\citep{sulisUsingMobilityData2018, bottaModellingUrbanVibrancy2021, wangMeasuringUrbanVibrancy2021}. Urban vibrancy is a concept that has been extensively theorised. Jane Jacobs was hugely influential in highlighting how urban design could encourage urban vibrancy and her arguments often focused on the maintenance and provision of social interactions in cities~\citep{jacobsDeathLifeGreat1961}. Her greatest addition to theory is an understanding that density and diversity in the physical structure of an urban place might affect its functional use ~\citep{moroniUrbanDensityJane2016} and that locations that are more diverse -- or more concentrated, in terms of their street networks, buildings, or `Points of Interest` -- may be the most vibrant locations. Thus, city planners should consider diversity and social accessibility because diversity provides social cohesion and supplies opportunities for spontaneous interactions, subsequently allowing high levels of creativity and activity that are accessible to the inhabitants and also maintain the community with a diverse socioeconomic background~\citep{perroneDowntownPeopleStreetlevel2019}.\par
We follow on from previous work by~\cite{bottaModellingUrbanVibrancy2021} that found that \emph{third places} -- i.e., places that humans use, that are neither home (\emph{first places}) or work (\emph{second places}) places and are specific in that they are used for social interactions -- are important predictors of urban vibrancy levels across age groups. Here, we study gender segregation and urban inequality through the lens of urban vibrancy. We explore the link between urban features and urban vibrancy and whether this differs for different genders, resulting in spatial segregation. We use a large data set that contains the presence of males and females in urban spaces, as measured via mobile phone activity data, as well as \emph{OpenStreetMap} geographical data, and residential census data. We use mobile phone activity data as a proxy measurement for urban vibrancy and analyse which urban features contribute to urban vibrancy for different social groups, particularly males and females. We find that there are differences between males and females in terms of urban vibrancy. Indeed, the differences relate to `Points of Interest` and transportation networks; however, there are both positive and negative spatial `spillovers` that exist across each city. We discuss how these differences could be accounted for in urban planning and design, and how human interaction with large technological systems provides a wealth of data that can complement that derived from more traditional methods of monitoring populations. This could allow social problems, such as spatial segregation, to be measured more accurately, and at faster rates, so that social problems might be solved more easily by policymakers and urban planners for the cities of the future. We aim to further the understanding of gender differences~\citep{vaitla_data_good} and segregation in urban spaces.

%% file: 03_data.tex
\section{Data}\label{sec:data}
We use three main data sources: (1) Italian census data that contains detailed information on where people live, (2) Call Detail Records data (CDR), derived from mobile phones, containing information on where people are at a high temporal granularity, and (3) \emph{OpenStreetMap}~\citep{osmOpenStreetMapContributors2017}. We use OSM data because it provides measurable features of urban environments. For each data type, we gather data for only the `metropolitan areas` of each city because these areas are the most densely populated areas of cities (see Supplementary Materials for summary statistics Table~\ref{tab:resident-information}). Metropolitan cities are areas that are linked to the city in terms of its culture and economy as well as its geographic proximity. Data for the metropolitan area boundary were gathered from the Italian Office for National Statistics (`ISTAT`) ~\citep{istatBordersAdministrativeUnits2023}. Data can be made available on request. We outline the data sources below.\par
\subsection{Italian census data}\label{subsec:italian-census-data}
The \emph{ISTAT} census~\citep{istatBordersAdministrativeUnits2023} is conducted every decade in Italy. Census sections are small, typically with 250 households, and provide total population and gender counts per section. We use 2011 census data downloaded from the ISTAT to confirm resident locations during key times of the day.\par
\subsection{Call Detail Record data}\label{subsec:call-detail-record-data}
We use mobile phone Call Detail Record (CDR) data from \emph{Gruppo TIM} (formerly \emph{Telecom Italia}) that was made available as part of their `\emph{Big Data Challenge 2015}` (described in~\cite{gruppotimDatasetSourceTIM2015}). The CDR data is available for seven different cities: \emph{Milan}, \emph{Rome}, \emph{Turin}, \emph{Naples}, \emph{Venice}, \emph{Palermo}, and \emph{Bari}, covering approximately two months (from 23:00 GMT on 2015-02-28 to 21:45 GMT on 2015-04-30). Each CDR data set has a corresponding grid designed by \emph{Telecom Italia}~\citep{gruppotimDatasetSourceTIM2015}. Each grid takes into account the topology of each city and the potential communication load. Each spatial grid was also designed to maintain the privacy of the inhabitants of each city. Subsequently, the grid polygons change shape in relation to the underlying mobile cells: cells typically get smaller the closer they are to the centre of each city. The activity of the data has a granularity of fifteen-minute intervals; however, there are time points in the CDR data that contain no records; this may occur because the number of users drops below three, and no data is recorded to preserve privacy, but there may also be further issues such as cutouts or problems in the collection of the data. The data contain the gender of the user who generated the CDR within the network  (see CDR summary statistics Table~\ref{tab:cdr-information}). The data contain a value for each gender and the gender is derived from the registration of the SIM card of male and female mobile phone users. Therefore, the data shows male and female use across time.\par
A related data set has already been used to understand how different age groups interact with cities~\citep{bottaModellingUrbanVibrancy2021}. This is possible because this type of data allows us to analyse and investigate the existence of differences across social groups instead of aggregating across a population, where information concerning the differences between social groups is neglected or reduced.\par
We utilize the mobile phone CDR data set as a vector-based proxy for measuring urban vibrancy in the cities under investigation because prior research findings demonstrate consistent effectiveness in studying urban vibrancy~\citep{sulisUsingMobilityData2018, wangMeasuringUrbanVibrancy2021, bottaModellingUrbanVibrancy2021}. Though graph-based metrics are also used to measure urban vibrancy~\citep{wangLearningUrbanCommunity2018a}, we approximate urban vibrancy in a vector-based way because the method allows fine temporal and spatial granularity and because we can separate the data according to gender. We intend to use these data to understand differences between social groups to understand how urban features contribute to a vibrant environment with respect to gender. We have aggregated the cell--user data across an array of time periods such that we get metrics pertaining to urban vibrancy in each grid cell.\par
\subsection{\emph{OpenStreetMap} data}\label{subsec:openstreetmap-data}
To understand cities, we must first create representations of their characteristics. Creating such representations has only recently been made easier thanks to collaborative projects such as \emph{OpenStreetMap}~\citep{osmOpenStreetMapContributors2017}. OSM is an open-source data repository generated and collected by volunteer collaborators. The data that is formed consists of large-scale geographic data that is made freely available to users. It is possible to download data on an array of city attributes including the networks, systems, and features of urban landscapes. Here, we retrieve data for each study area; however, it is important to note that the data that was downloaded was the most up-to-date version of the urban features. What we derive from these data is explained systematically in the next section.

%% file: 04_methods.tex
\section{Methods}\label{sec:methods}
\subsection{A proxy for urban vibrancy: Call Detail Records}\label{subsec:a-proxy-for-urban-vibrancy:-call-detail-records}
As a proxy measure for urban vibrancy, we use CDR data as it indicates the presence of inhabitants throughout the day. We calculate gender differences by subtracting the male value from the female value:
\begin{linenomath}
    \begin{align}
        \Delta_{i} = M_{i} - F_{i}\label{eq:difference}
    \end{align}
\end{linenomath}
where \(\Delta_{i}\) represents the vector of differences, \(M_{i}\) is the vector of male users, and \(F_{i}\) is the female users.\par
Gender identity was only available for users who disclosed it when acquiring their SIM cards. Users who did not disclose their gender were excluded from the analysis. As described in Section~\ref{subsec:call-detail-record-data}, gaps in the data occur when the number of users falls below three to protect their anonymity. To fill in these gaps, we assumed a value of zero for these time points. The grid consists of cells of varying sizes, so we normalized the raw data by the area of each grid cell to obtain a population density that accounts for the usage and topology of each cell in relation to the city.
\subsection{Cities as networks: Independent variables}\label{subsec:cities-as-networks-independent-variables}
We downloaded for each static grid cell of each city a range of features shown to be related to urban vibrancy by previous research~\citep{sungEvidenceJacobsStreet2013, bottaModellingUrbanVibrancy2021, yuIntergenerationalDifferencesUrban2022a, chenInvestigatingSpatiotemporalPattern2022a}. We processed these features in two ways as outlined below.\par
\subsubsection{\emph{Density} in urban features}\label{subsubsec:density-in-features}\hfill\par
By Jacobs~\citep{jacobsDeathLifeGreat1961}, increased feature density was a promoting factor in urban vibrancy because of the increased activity. The density of buildings, highways, networks, intersections, or `Points of Interest` in a place, all have the potential to provide more opportunities for activities because of the increased number of users of those locations, or the increased vehicular or pedestrian access; however, importantly, these might differ across genders. Here, we define density as the concentration of a feature type within a given area. To arrive at the density value, we first download features from the free geographical database: \emph{OpenStreetMap} (OSM; see Section~\ref{sec:data}). We construct feature collections of the buildings, transport networks, and `Points of Interest` found in each cell of each city. We took the total count per cell and divided it by the total area of that cell to give a value of the feature density. For the networks, we calculated the average length of transport networks or the average number of intersections, that are accessible for (1) pedestrians, (2) cyclists, and (3) drivers. We used the following calculation to determine feature density:
\begin{linenomath}
    \begin{align}
        \rho = \frac{N}{A}\label{eq:density}
    \end{align}
\end{linenomath}
where \(\rho\) is the density, \(N\) is the total number of features in the geometry, and \(A\) is the area of the geometry. The values were added to the grid cells for each city (see Figure~\ref{fig:si_2}).\par
\subsubsection{\emph{Diversity} in urban features}\label{subsubsec:Diversity-in-urban-features}\hfill\par
According to Jacobs, it is the diversity in features that increases and encourages a location's vibrancy~\citep{jacobsDeathLifeGreat1961}. Similarly to above, in a place, the diversity of buildings, highways, or `Points of Interest` all have the potential to provide more opportunities for activities because of the increased usage of those locations; however, these might differ across genders. To gather data on urban feature diversity we use the same downloaded OSM feature collections. We use the Shannon-Wiener diversity index~\citep{shannonMathematicalTheoryCommunication1948} to calculate the diversity of features. The diversity index is calculated as follows:
\begin{linenomath}
	\begin{align}
        H=-\sum_{i=1}^{M} P_i\,log_2\,P_i\label{eq:shannon}
	\end{align}
\end{linenomath}
where \(H\) is the diversity index, \(M\) is the total number of categories in the geometry feature, and \(P_i\) is the frequency of the \(i\textsuperscript{th}\) category. The values were added to the grid cells for each city (see Figure~\ref{fig:si_2}). We detail the `Points of Interest` variables below.\par 
\paragraph{`Points of Interest`:}\label{par:points-of-interest}`Points of Interest` are important features that directly relate to urban vibrancy. We collected all `Points of Interest` from OSM found under the amenity, building, leisure, shop, and sport tags in the OSM database to construct a `Points of Interest` collection for each city. We manually labeled the points using the same label collection as~\cite{moroMobilityPatternsAre2021b} and~\cite{fanDiversityDensityExperienced2022}, based on the Foursquare classification system, which includes 14 categories. This taxonomy of labels is as follows: (1) \emph{Arts / Museum}, (2) \emph{City / Outdoors}, (3) \emph{Coffee / Tea}, (4) \emph{College}, (5) \emph{Entertainment}, (6) \emph{Food}, (7) \emph{Grocery}, (8) \emph{Health}, (9) \emph{Residential}, (10) \emph{Service}, (11) \emph{Shopping}, (12) \emph{Sports}, (13) \emph{Transportation}, and (14) \emph{Work}. We considered these labels because they represent the most frequently visited locations and are likely to be important for segregation~\citep{moroMobilityPatternsAre2021b}.
\paragraph{`Third Places`:}\label{par:third-places} We constructed a collection of `\emph{Third Places}`~\citep{oldenburgThirdPlace1982b}. Third Places are locations that are neither home nor workplaces and are considered to be vitally important in terms of urban vibrancy because they allow for impromptu everyday gatherings in urban locations that result in positive effects for communities~\citep{jeffresImpactThirdPlaces2009a, bottaModellingUrbanVibrancy2021} and because people spend a significant fraction of their free time in third places. For this analysis, we considered that there may be differences between genders in how they use amenities like shops, pubs, cafés, or community centres. And that this was likely to be used differently depending on general discrepancies in urban mobility diversities and, amongst others, related to socioeconomic characteristics.\\
Here, using the same `Points of Interest` tags used in OSM (i.e., `amenity`, `building`, `leisure`, `shop`, and `sport`), we calculated density and diversity for third places (as defined in Section~\ref{subsubsec:density-in-features} and~\ref{subsubsec:Diversity-in-urban-features}) across all grid cells and cities. We manually labeled third places based on~\cite{jeffresImpactThirdPlaces2009a}'s categorization of (1) \emph{eating and drinking}, (2) \emph{organized activities}, (3) \emph{outdoor}, and (4) \emph{commercial venues}, and added a fifth label of \emph{commercial services}. Commercial services are based on the locations where people might receive a service or go with the intent of buying something, but where you may also have opportunities for social interactions that might be brief compared to the other groupings. We included this label to capture the potential for brief social interactions in locations like banks and pharmacies that would have otherwise been removed due to~\cite{jeffresImpactThirdPlaces2009a} definition. Only those locations that fit within these categories were considered third places.\par

%% file: 04_statistics.tex
\subsection{Statistical approach}\label{subsec:statistical-approach}
In this analysis, the Call Detail Record (CDR) data and census data have different spatial grids. We use the geometry of CDR data as our main reference and extract OSM data for each cell. We interpolate census data to the same spatial grid using areal interpolation~\citep{comberSpatialInterpolationUsing2019a, bergroth24hourPopulationDistribution2022a} to enable correlation analysis and spatial linear regression at the grid cell level (for an overview of the methodology, see~\ref{fig:si_1}).\par
We perform a correlation analysis to test the data's representativeness by comparing the CDR data (we used the nighttime values only) with census counts that have been converted to density estimates, matching the CDR data in terms of spatial scale and intensive property. We compare both male and female nighttime CDR values with their respective interpolated census data and use Kendall's correlation coefficient as it is distribution-free and more suitable for spatial data~\citep{hamedDistributionKendallTau2011}. We carry out this analysis in all cities.\par
We aim to model male-female differences for each city while also building an aggregated model to identify common trends. We refer to this aggregated model in many of the analyses below for clarity. To ensure comparability, we standardize all variables. We then construct an ordinary least squares (OLS) model as a baseline method for estimating the regression \(\beta\) coefficients and evaluating the importance of spatial extensions to OLS. We use the following linear function to explain male-female differences as a function of a set of separate features denoted by \(X\) (See Section~\ref{subsec:a-proxy-for-urban-vibrancy:-call-detail-records}):
\begin{linenomath}
    \begin{align}
    Y = \beta_{i} X_{i} + \epsilon\label{eq:non-spatial-regression}
    \end{align}
\end{linenomath}
where \(Y\) represents the male-female differences as a response variable consisting of a value proportional to the users per gender used here as a proxy for activity and a measure of vibrancy (see Section~\ref{subsec:a-proxy-for-urban-vibrancy:-call-detail-records}), \(\beta_{i}\) are the regression coefficients, \(X_{i}\) is the independent variables, and \(\epsilon\) is the error.\par
Our data may have spatial autocorrelation, which violates some assumptions of a basic regression model. We address this issue in the following sections.\par
We diagnosed spatial dependence by analyzing OLS model residuals. The error terms may not be independent due to a spatial relationship, so we checked for spatial structure and the need for spatial models using Moran's \emph{I} analyses from~\cite{wardSpatialRegressionModels2019}. Spatial clustering implies spatial dependence, and so requires spatial models.\par
We use maximum likelihood estimation to create spatial lag and spatial error models~\citep{anselinGeoDaIntroductionSpatial2006, wardSpatialRegressionModels2019}. The Lagrange multiplier and AIC difference from OLS were calculated. We utilized \emph{Queen's contiguity} based on grid-cell geometry for spatial weights' matrix, connecting centroids of observations to those with shared vertices~\citep{reyPySALPythonLibrary2010}. We used row transformation to normalize the weights' values to achieve an average of variable values in each observation's neighbourhood. We chose Queen's contiguity due to the irregular grid size. To assess the sensitivity of the choice of spatial weights, we replicate the same analysis using both Rook's contiguity and k-nearest neighbours' distance as weighting schemes, and we find qualitatively similar results (see~\ref{fig:si_3}).\par
We consider the spatial error model (SEM) as our first model, which incorporates spatial dependence using a spatially lagged error term:
\begin{linenomath}
    \begin{align}
        Y = X_{i} \beta_{i} + u, u = \rho W u + \epsilon \label{eq:sem}
    \end{align}
\end{linenomath}
where \(Y\) is the male-female differences as a response variable, \(X_{i}\) represents the explanatory variables, \(\beta_{i}\) are the regression coefficients, \(u\) is the first error term, \(\rho\) is a scalar of the spatial lag parameter, \(W\) is the weights' matrix (Queen's contiguity), and \(\epsilon\) is the spatially independent error term.\\
We also utilized spatial lag models (SAR) where dependent variables were spatially lagged, providing coefficients for both \emph{direct} and \emph{indirect} effects of independent variables on the response and mean activity of neighbouring grid cells~\citep{lesageSpatialGrowthRegressions2008}. The SAR model terms are as follows:
\begin{linenomath}
    \begin{align}
        Y = X_{i} \beta_{i} + \rho WY u + \epsilon \label{eq:sar}
    \end{align}
\end{linenomath}
where \(Y\) is the vector of the response variable, \(X\) represents the explanatory variables, \(\beta\) is the regression coefficients, \(\rho\) is a scalar of the spatial lag parameter, \(WY\) is the weights' matrix (Queen's contiguity), and \(\epsilon\) is the spatially independent error term.\par
For each model, we calculated the direct effect, indirect effects, and total effects due to the challenges associated with interpreting predictors unit change under `spillover' effects~\citep{anselinModernSpatialEconometrics2014}. Spatial spillover effects are indirect effects and refer to secondary impacts that result from the direct effects. To compute these values, we first derived estimated coefficients for exogenous variables in the model, which yielded the direct effect that shows the influence of one spatial unit on another~\citep{lesageSpatialGrowthRegressions2008}. Direct effects are calculated as:
\begin{linenomath}
    \begin{align}
        DE = \beta
    \end{align}
\end{linenomath}
where \(DE\) are the direct effects, \(\beta\) are the coefficients of the SAR model without the spatial lag term.\par
Contained within the direct effects are both the indirect effects and the total effects~\citep{guomengAnalysisDirectEffect2021}. To split these apart, we extract the spatial lag term from the coefficients and divide the coefficients by 1 minus the spatial lag term multiplied by the largest eigenvalue of the spatial weights' matrix which effectively normalizes the coefficients~\citep{lesageSpatialGrowthRegressions2008, bivandComparingImplementationsEstimation2015}. Total effects take into account the full range of impacts that a particular change may have on the spatial system as a whole~\citep{lesageSpatialGrowthRegressions2008}. Total effects are calculated as:
\begin{linenomath}
    \begin{align}
        TE = \beta / (1 - \rho * \lambda)
    \end{align}
\end{linenomath}
where \(TE\) are the total effects, \(\beta\) are the coefficients of the exogenous variable in the spatial lag model, \(\rho\) is the spatial lag term, and \(\lambda\) is the maximum eigenvalue from the spatial weights' matrix.\par
Finally, we calculate the indirect effects. These refer to the secondary impacts that result from the direct effects. Indirect effects are a measure of any spillover effects that occur beyond the immediate spatial units~\citep{lesageSpatialGrowthRegressions2008}. Indirect effects are calculated as:
\begin{linenomath}
    \begin{align}
        IE = DE - TE
    \end{align}
\end{linenomath}
where \(IE\) are the indirect effects, \(DE\) are the direct effect and \(TE\) are the total effects.\par
Using the above methods and data, we created a hierarchy of models that aggregated CDR data at different time periods. These included: (i) All daytime data (08:00-20:00), (ii) Weekdays (Monday-Thursday) versus Weekends (Friday-Sunday) within the all-day time period, and (iii) Twenty-four-hour day data divided into Morning (06:00-12:00), Afternoon (12:00-18:00), Evening (18:00-00:00), and Night (00:00-06:00) categories. Models were run for individual cities and for all cities combined. This approach enabled us to study how vibrancy relates to gender and urban features at different times and investigate variations over time.

%% file: 05_results.tex
\section{Results}\label{sec:results}
First, we use Kendall’s rank correlation coefficient~\citep{kendallNewMeasureRank1938} to compare the census and the Call Detail Record (CDR) data to check representativeness. Across all cities, we find results were positive and significant at the $5 \%$ level (see Figure~\ref{fig:figure-si_5}). The lowest \(\tau\) for females is 0.55 (Bari) whereas the highest \(\tau\) for females is 0.72 (Torino). For males, the lowest \(\tau\) is 0.56 (Bari) whereas the highest \(\tau\) for males is 0.7 (Torino). These strong correlations between the CDR--when selecting only the nighttime values--and census data show that the CDR data are broadly representative of the census data. This suggests that the values of the daytime should also be representative of the movements of the general population.\par
Next, we use Kendall's rank correlation coefficient to compare each urban feature with the male-female differences. We adjust our results using false discovery-rated detection to correct for the rate of type I errors in the null hypothesis. We found that, except for a few instances, all of our results were positive and were highly significant at the $5\%$ level (\emph{P} \(<\) 0.05). Also, between density and diversity metrics, diversity often had a smaller association contrasting with previous work on age groups~\citep{bottaModellingUrbanVibrancy2021}, we found that smaller cities had larger associations with male-female differences. See Table~\ref{tab:correlation-analysis} for the full correlation analysis results.\par
\begin{table*}[htp]
\small
\setlength\tabcolsep{.5pt}
\setlength{\arrayrulewidth}{1pt}
\caption{Correlation between male-female differences and all urban features. For each city and the aggregation of all cities, we calculate the correlation coefficient and associated $p$-value. We have separated the variables into categories of density and diversity. We use Kendall's tau (\(\tau\)) rank correlation coefficient because of its particular suitability for spatial data~\citep{hamedDistributionKendallTau2011}. We can see both the associational strength and directionality of each correlation. We also calculated the average for each variable based on each city but not the aggregation of all cities. $p$-values are shown using their significance stars, where [0 - 0.001] is '***', [0.001 - 0.01] is '**', [0.01 - 0.05] is '*', [0.05 - 0.1] is '.', and [0.1 - 1.0] has no symbol. Cities are ordered by the number of grid cells.\hfill \break}\label{tab:correlation-analysis}
\begin{tabular}[t]{lrrrrrrrrrrrrrrrrrr}
\multicolumn{1}{c}{Variable} & \multicolumn{2}{c}{Aggregated} & \multicolumn{2}{c}{Milano} & \multicolumn{2}{c}{Roma} & \multicolumn{2}{c}{Torino} & \multicolumn{2}{c}{Napoli} & \multicolumn{2}{c}{Venezia} & \multicolumn{2}{c}{Palermo} & \multicolumn{2}{c}{Bari} & \multicolumn{1}{c}{Average} \\
\cmidrule(l{3pt}r{3pt}){1-1} \cmidrule(l{3pt}r{3pt}){2-3} \cmidrule(l{3pt}r{3pt}){4-5} \cmidrule(l{3pt}r{3pt}){6-7} \cmidrule(l{3pt}r{3pt}){8-9} \cmidrule(l{3pt}r{3pt}){10-11} \cmidrule(l{3pt}r{3pt}){12-13} \cmidrule(l{3pt}r{3pt}){14-15} \cmidrule(l{3pt}r{3pt}){16-17} \cmidrule(l{3pt}r{3pt}){18-18}
 & \(\tau\) & $p$ & \(\tau\) & $p$ & \(\tau\) & $p$ & \(\tau\) & $p$ & \(\tau\) & $p$ & \(\tau\) & $p$ & \(\tau\) & $p$ & \(\tau\) & $p$ & \\
\midrule
\addlinespace[0.3em]
\multicolumn{19}{l}{\textbf{Density}}\\
\hspace{1em}\texttt{Building} & 0.41 & *** & 0.25 & *** & 0.49 & *** & 0.42 & *** & 0.40 & *** & 0.60 & *** & 0.64 & *** & 0.62 & ***  & 0.53\\
\hspace{1em}\texttt{Road network length} & 0.37 & *** & 0.19 & *** & 0.46 & *** & 0.48 & *** & 0.41 & *** & 0.51 & *** & 0.58 & *** & 0.51 & ***  & 0.49\\
\hspace{1em}\texttt{Cycling network length} & 0.36 & *** & 0.21 & *** & 0.52 & *** & 0.47 & *** & 0.43 & *** & 0.53 & *** & 0.53 & *** & 0.46 & ***  & 0.49\\
\hspace{1em}\texttt{Walking network length} & 0.36 & *** & 0.27 & *** & 0.55 & *** & 0.48 & *** & 0.45 & *** & 0.53 & *** & 0.55 & *** & 0.44 & ***  & 0.5\\
\hspace{1em}\texttt{Highway} & 0.41 & *** & 0.34 & *** & 0.65 & *** & 0.57 & *** & 0.65 & *** & 0.74 & *** & 0.71 & *** & 0.62 & ***  & 0.66\\
\hspace{1em}\texttt{Road network node} & 0.40 & *** & 0.20 & *** & 0.51 & *** & 0.52 & *** & 0.51 & *** & 0.54 & *** & 0.62 & *** & 0.55 & ***  & 0.54\\
\hspace{1em}\texttt{Cycling network node} & 0.39 & *** & 0.25 & *** & 0.55 & *** & 0.52 & *** & 0.53 & *** & 0.56 & *** & 0.62 & *** & 0.54 & ***  & 0.56\\
\hspace{1em}\texttt{Walking network node} & 0.38 & *** & 0.29 & *** & 0.58 & *** & 0.52 & *** & 0.51 & *** & 0.57 & *** & 0.63 & *** & 0.53 & ***  & 0.56\\
\hspace{1em}\texttt{Point of interest} & 0.42 & *** & 0.29 & *** & 0.58 & *** & 0.50 & *** & 0.13 & * & 0.65 & *** & 0.61 & *** & 0.66 & ***  & 0.52\\
\hspace{1em}\texttt{Third place} & 0.36 & *** & 0.28 & *** & 0.55 & *** & 0.51 & *** & 0.45 & *** & 0.65 & *** & 0.61 & *** & 0.49 & ***  & 0.54\\
\addlinespace[0.3em]
\multicolumn{19}{l}{\textbf{Diversity}}\\
\hspace{1em}\texttt{Building} & 0.08 & *** & 0.14 & *** & 0.29 & *** & 0.32 & *** & -0.04 &  & -0.23 & ** & -0.16 & . & 0.42 & ***  & 0.1\\
\hspace{1em}\texttt{Highway} & 0.00 &  & -0.10 & ** & 0.19 & *** & 0.02 &  & 0.01 &  & 0.04 &  & -0.02 &  & 0.35 & ***  & 0.1\\
\hspace{1em}\texttt{Point of interest} & 0.10 & *** & 0.11 & ** & 0.32 & *** & 0.03 &  & 0.44 & *** & 0.44 & *** & 0.13 &  & 0.10 &   & 0.24\\
\hspace{1em}\texttt{Third place} & 0.06 & ** & -0.04 &  & 0.27 & *** & -0.04 &  & 0.06 &  & 0.22 & ** & 0.12 &  & 0.20 & *  & 0.14\\
\hspace{1em}\texttt{Commercial venue (third place)} & 0.30 & *** & 0.27 & *** & 0.42 & *** & 0.46 & *** & 0.27 & *** & 0.54 & *** & 0.48 & *** & 0.34 & ***  & 0.42\\
\hspace{1em}\texttt{Organised activity (third place)} & 0.20 & *** & 0.10 & ** & 0.40 & *** & 0.15 & *** & 0.20 & ** & 0.46 & *** & 0.22 & * & 0.46 & ***  & 0.31\\
\hspace{1em}\texttt{Outdoor (third place)} & 0.26 & *** & 0.17 & *** & 0.46 & *** & 0.38 & *** & 0.29 & *** & 0.57 & *** & 0.32 & *** & 0.31 & ***  & 0.39\\
\hspace{1em}\texttt{Eating drinking (third place)} & 0.31 & *** & 0.26 & *** & 0.50 & *** & 0.41 & *** & 0.22 & *** & 0.53 & *** & 0.36 & *** & 0.33 & ***  & 0.39\\
\hspace{1em}\texttt{Commercial service (third place)} & 0.27 & *** & 0.24 & *** & 0.45 & *** & 0.40 & *** & 0.24 & *** & 0.46 & *** & 0.42 & *** & 0.37 & ***  & 0.39\\
\end{tabular}
\end{table*}
We use Moran's spatial autocorrelation analysis to determine the global Moran's \emph{I} of the data and to further understand the data in terms of its spatial dependence. We extract residuals of an ordinary least squares (OLS) model at each level of the model hierarchy. We calculate Moran's \emph{I} statistics for the male-female differences in each city. For clarity, we report only the minimum and maximum in the lowest level of the hierarchy (this does not include the aggregated model which is excluded from this analysis due to its obvious spatial clustering in different cities), i.e.\ the daytime data (08:00-20:00; see Section~\ref{subsec:statistical-approach}). The lowest value for Moran's \emph{I} was 0.013, the maximum was 0.8, and the mean was 0.2. Of the seven cities in the analysis, two were non-significant at the $5\%$ level, these were Bari and Napoli; all the rest were significant (\emph{P} \(<\) 0.05). These values confirm the presence of spatial clustering and spatial dependence; however, these values alone cannot provide a full account of the presence or absence of spatial clustering or fully understand the spatial structure of the data; this, however, provides evidence that spatial models may be more appropriate. See Supplementary Materials Figure~\ref{fig:si_4} and Table~\ref{tab:moran-i} for full Moran's \emph{I} analysis results.\par
To further test the presence of spatial clustering, we calculate the Lagrange multiplier test statistic (both non-robust and robust) for the OLS model of each city and at each level of the analysis. We did this to identify the type of model most suitable for the analysis. This provides a measure to inform us whether to use a simple linear model or use either the spatial lag models (SAR) or the spatial error models (SER)~\citep{lesageIntroductionSpatialEconometrics2009a}. At the same time, we calculate the Akaike information criterion (AIC) of each of the models. We find that the SAR models most often contain the highest values (Minimum LMerr = 0.062, \emph{P} is NS, Maximum LMerr = 410.104; Minimum LMlag = 9.56, \emph{P} \(<\) 0.001, Maximum LMlag = 539.642). The same relationship is also found for the robust methods. The differences between the AIC of the OLS and the AIC of the spatial models were consistently greater for the SAR models. Because these tests consistently pointed toward using the SAR models, and we observe spatial clustering, we continue the rest of our analysis with SAR models.\par
\begin{figure*}[t]
	\includegraphics[width=\linewidth]{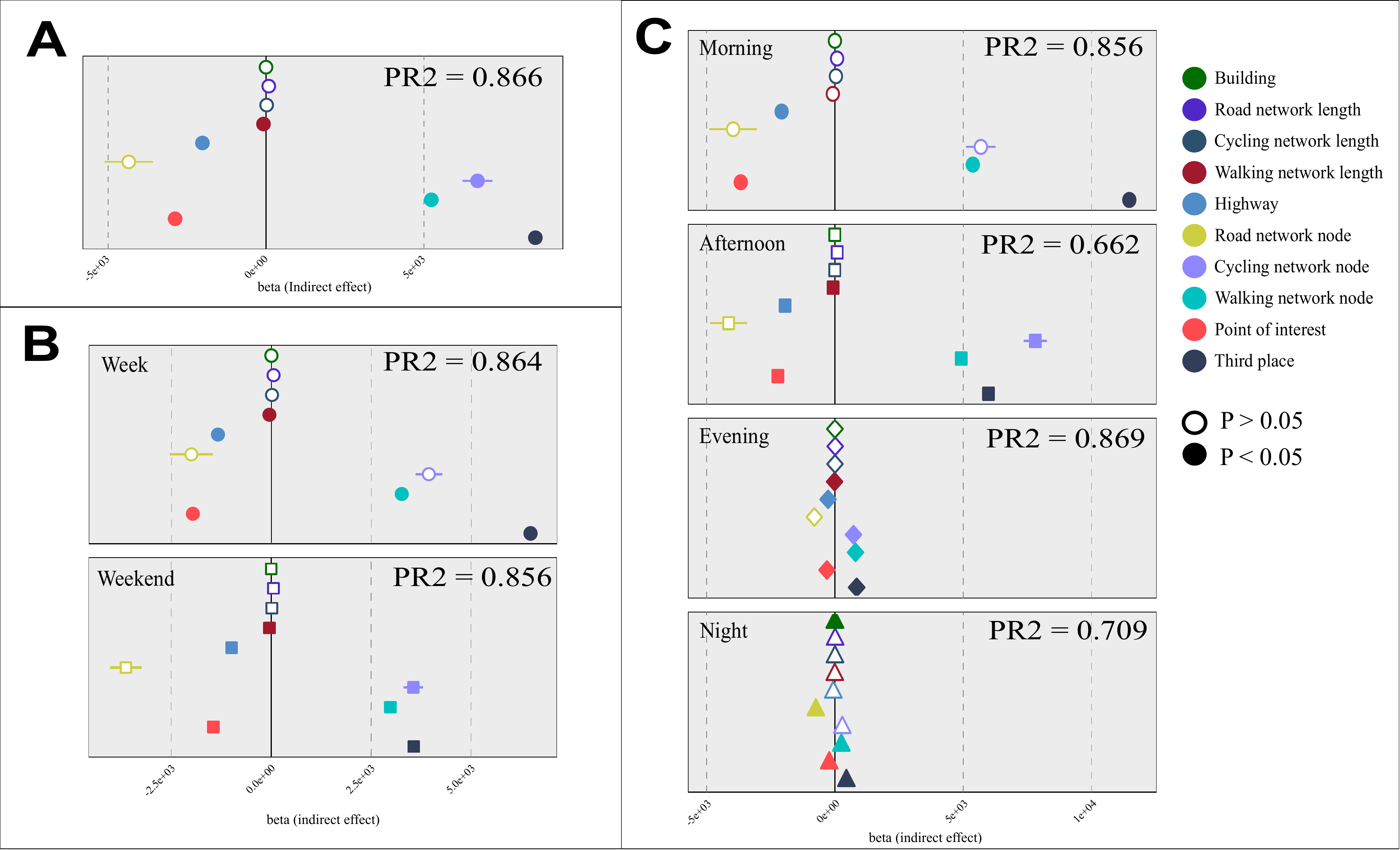}
	\caption{The relationship between density in features and male-female differences for the spatial model aggregating all cities. The plot shows the significance of the direct effect; however, the \(\beta\) coefficients represent the indirect effect (see~\ref{subsec:statistical-approach} for definitions). The plot displays all density variables, with the y-axis showing the variables for each model and the colour representing each variable. Panel (A) shows all daytime data between 08:00 and 20:00; Panel (B) displays weekday (Monday-Thursday) versus weekend (Friday-Sunday) data within the same time period; Panel (C) shows data averaged into four categories: Morning (06:00-12:00), afternoon (12:00-18:00), Evening (18:00-00:00), and Night (00:00-06:00). To aid comparison, shapes denote categories in Panels (B) and (C). Significance is indicated by closed and open shapes (\emph{P} \(<\) 0.05 and \emph{P} \(>\) 0.05, respectively), and each shape shows the error bars as a horizontal line. The pseudo-r-squared is reported for each panel.}\label{fig:figure-main_fig_1}
\end{figure*}

We find fairly consistent results across this hierarchy. Firstly, we find that smaller cities had larger amounts of error in the estimates than larger cities and with more variation in the coefficients, most likely due to the size and number of cells in the smaller cities. Secondly, there are no strong gender differences between the night and day (see Figures~\ref{fig:figure-main_fig_1},~\ref{fig:si_6} and~\ref{fig:si_7}). In the larger cities, we find a significant positive indirect effect between male-female differences and third-place density (see Figure~\ref{fig:figure-main_fig_1}); this relationship is consistent across the three largest cities and the aggregated model. We also found a significant negative indirect effect between male-female differences and the density of `Points of Interest` (see Figure~\ref{fig:figure-main_fig_1}). The pattern was again similar across, this time, the four largest cities and the aggregated model. We also find a relationship between highway density and all three of the intersection variables; however, they did not share the direction. Both cycling and walking intersection density were significant with positive indirect effects; however, for road intersections, though there was a negative indirect effect, this was not significant. The highway density was only significant in the aggregate model (see Figure~\ref{fig:figure-main_fig_1}). We found that diversity metrics generally were not significant; however, we found that the `organised activity` and `outdoor` third place categories were significant; this was not the case for third places generally (see Figures~\ref{fig:si_6} and~\ref{fig:si_7}).\\
We find that the strongest effect, which was also the most consistent, was the positive indirect effect of the third place density. We consider this an interesting finding when coupled with the significant negative indirect effect found in the `Points of Interest` variable because the directionality of the `Points of Interest` variable -- without accounting for the social aspect of the third places -- are opposing one another. This could suggest that locations that fall under our five categories for third places are not equally used by each gender whereas, the `Points of Interest` as a whole are used more equally.\par
For each level of the nested hierarchy, we report the results of the pseudo-r-squared values from the models. The pseudo-r-squared is the squared correlation between the dependent variable and the predictions of the dependent variable~\citep{anselinSpatialEconometricsMethods1988}. These values are a measure of goodness-of-fit and are used to understand the relationship between the model variables. Our models exhibited relatively high values consistently across cities and across the hierarchy of models highlighting a good relationship between independent and dependent variables (see Figures~\ref{fig:figure-main_fig_1},~\ref{fig:si_6}, and~\ref{fig:si_7}).\par

%% file: 06_discussion.tex
\section{Discussion}\label{sec:discussion}
In this study, we have focused on modelling \emph{urban vibrancy}--a measure of the dynamic activity of human beings in urban environments. For this, we have considered seven of the largest cities in Italy. We asked how urban features might contribute to a vibrant environment and how they might vary across social groups, especially concerning gender. We hypothesised that there would be differences for different genders because, firstly, heterogeneity exists generally in how people interact with urban environments~\citep{depalmaHeterogeneityStatesUrban1988} but, secondly, that similarities might correlate most closely with groups such as gender due to similarity in socioeconomic characteristics or general behaviours. We used a computational approach to reveal any potential socio-spatial segregation across our study areas, and we used a range of relevant urban features taken from urban vibrancy theory~\citep{sungEvidenceJacobsStreet2013, bottaModellingUrbanVibrancy2021, yuIntergenerationalDifferencesUrban2022a, chenInvestigatingSpatiotemporalPattern2022a}. To model urban vibrancy, we used data showing the presence of mobile phone users as a proxy--another established methodology~\citep{jiaMeasuringVibrancyUrban2019, bottaModellingUrbanVibrancy2021}.\par
We have uncovered a variety of findings. First, we have been able to study urban vibrancy--and potential segregation in urban vibrancy--by using high-frequency Call Detail Record (CDR) data and open-source geographical data. We show that it is possible to do this with high predictive power and goodness-of-fit, and we do this across a model hierarchy that accounts for movement behaviours in order to reflect the reality of life in cities. Second, we have furthered discussions from previous works that focused on the importance of third places in cities: we found significant evidence that an increase in the density of third places in a given area is associated with larger male-female differences in urban vibrancy whereas an increase in the density of `Points of Interest` overall (i.e., without specifically considering third places) is associated with smaller male-female differences. The evidence that third places are associated with larger differences could suggest that locations that we have defined as third places, i.e., locations that fall under our five-category system (see Section~\ref{sec:methods}), are not equally used by each gender. This evidence does not necessarily mean that increases in the density of third places increase differences; however, it may be that certain types of third places are unequally used across genders. Reasons for this could be based on cultural differences or socioeconomic factors. Another part to consider is the clustered nature of third places in cities: a positive indirect effect indicates increases in male-female differences but also that the variable is positively correlated with the neighbouring values of male-female differences. It is important to consider that third places are places that are likely to cluster geographically with other factors such as economic activity or environmental conditions. More evidence would be needed to uncover further details but a similar methodology could be used with extensions and additional analyses. One such methodological adaptation could be the use of a Geographically Weighted Regression. This would help to understand the predictive power across cities whilst exploring potential spatial biases in the data.\par
Within our analysis, we can identify a number of limitations, and it is important to acknowledge these and discuss them here. Firstly, our CDR data is used as a proxy for urban vibrancy measurement; these data are from Telecom Italia, just one provider. Though this is the largest provider in Italy, these data do not capture the entire population and so may contain unknown biases. A full account of the general population could be gained by using multiple providers and may improve our overall analysis. It is also the case that the data were only derived from phone calls; this clearly misses a breadth of other communication methods and could potentially hide biases in the data due to the myriad of different ways people communicate today. Furthermore, gender information is derived from SIM purchases, but this is likely to not be an exact representation of the gender of users. However, we also note that our validation with the census data shows a good correlation with the mobile phone data, suggesting that these issues may be relatively limited (see Figure~\ref{fig:figure-si_5}). A second potential problem is that the data are from differing time periods: we have taken census data from 2011, CDR data from 2015, and \emph{OpenStreetMap} data from 2022. This undoubtedly introduces some biases in the analysis; however, we expect them to be small and not affect the overall results.\par\par
In this study, we have considered the modelling of \emph{urban vibrancy} with respect to gender differences. We found that the density of different collections of `Points of Interest` are simultaneously associated with both decreases and increases in male-female differences. This was the case when we gave a broad-scale use-category (`Points of Interest`) and a fine-scale use-category that considers the social context of a place (third place). This adds further evidence for the importance of characterising third places when studying urban environments and urban vibrancy. This evidence also suggests that comparing different collections of `Points of Interest` could hold interesting avenues for further research relating to urban vibrancy. We have shown that this could also provide details on the potential segregation we find existing in cities today.\par
To conclude, our analysis provides further evidence and support for the use of CDR and crowdsourced data to understand large-scale movement behaviours and how we can use these data to understand the social fabric of urban life. In turn, this could provide evidence for the design of our future urban environments.\par

%% file: 07_supp.tex
\clearpage
\section*{Supplemental material}\label{sec:supplemental-materiall}
\begin{table*}[htbp!]
\caption[Resident population]{Summary statistics for each metropolitan study area. The table shows (i) the total resident population for both males and females, (ii) the total area of each study area, and (iii) the total grid cells taken from~\cite{gruppotimDatasetSourceTIM2015}. 2015 census data taken from~\cite{istatBordersAdministrativeUnits2023}.}\label{tab:resident-information}
\centering
\setlength\tabcolsep{2pt}
\begin{tabular}{llllll}
    \toprule
    City name &  Males        &  Females     &  Total         &  Grid cell \\
              &  (thousands)  &  (thousands) &  area ($km^2$) &  count\\
    \midrule
    \texttt{Milano}   &  1218 &  1334 &  1128  &  794\\
    \texttt{Roma}     &  1477 &  1636  &  3746  &  739\\
    \texttt{Torino}   &  893  &  966  &  2850  &  494 \\
    \texttt{Napoli}   &  800  &  846   &  643   &  267\\
    \texttt{Venezia}  &  231  &  251  &  1200  &  157 \\
    \texttt{Palermo}  &  293  &  317  &  2294  &  130\\
    \texttt{Bari}     &  309  &  320 &  2363  &  110\\
    \bottomrule
\end{tabular}
\end{table*}
\begin{table*}[htp]
\caption [CDR population]{Mean activity across the day. Values are proportional to the number of males and females as an average (mean) across the day and each city. Values are derived from the Call Detail Record data taken for each metropolitan study area. The table shows the values for both males and females in each period of the day. The periods are (i) Morning (06:00-12:00), (ii) Afternoon (12:00-18:00), (iii) Evening (18:00-00:00), and (iv) Night (00:00-06:00). Data are taken from~\cite{gruppotimDatasetSourceTIM2015}. Cities are ordered by the total number of grid cells.}\label{tab:cdr-information}
\centering
\setlength\tabcolsep{2pt}
\begin{tabular}{lrrrrrrrr}
\toprule
City name & Gender  & Morning & Afternoon & Evening & Night \\
\midrule
\texttt{Milano} & F   &   9.509 &    10.460 &   1.991 & 0.644 \\
        & M   &    12.464 &    12.942 &   2.221 & 0.880 \\
\addlinespace
\texttt{Roma} & F   &    17.244 &    17.325 &   3.654 & 1.183 \\
        & M   &    23.020 &    21.585 &   4.134 & 1.650 \\
\addlinespace
\texttt{Torino} & F   &    9.373 &    10.091 &   1.941 & 0.610 \\
        & M   &    12.190 &    12.558 &   2.106 & 0.834 \\
\addlinespace
\texttt{Napoli} & F   &    51.808 &    51.584 &  18.483 & 4.382 \\
        & M   &    78.761 &    73.633 &  22.874 & 7.067 \\
\addlinespace
\texttt{Venezia} & F   &    3.503 &     3.783 &   0.498 & 0.196 \\
        & M   &    6.562 &     6.569 &   0.785 & 0.407 \\
\addlinespace
\texttt{Palermo} & F   &    29.645 &    27.138 &   6.924 & 2.185 \\
        & M   &    49.927 &    43.120 &   9.590 & 4.072 \\
\addlinespace
\texttt{Bari} & F   &    21.804 &    20.271 &   5.867 & 1.878 \\
        & M   &    38.773 &    33.843 &   8.408 & 3.745 \\
\bottomrule
\end{tabular}
\end{table*}
\begin{figure*}[htp]
    \centering
	\includegraphics[scale=1]{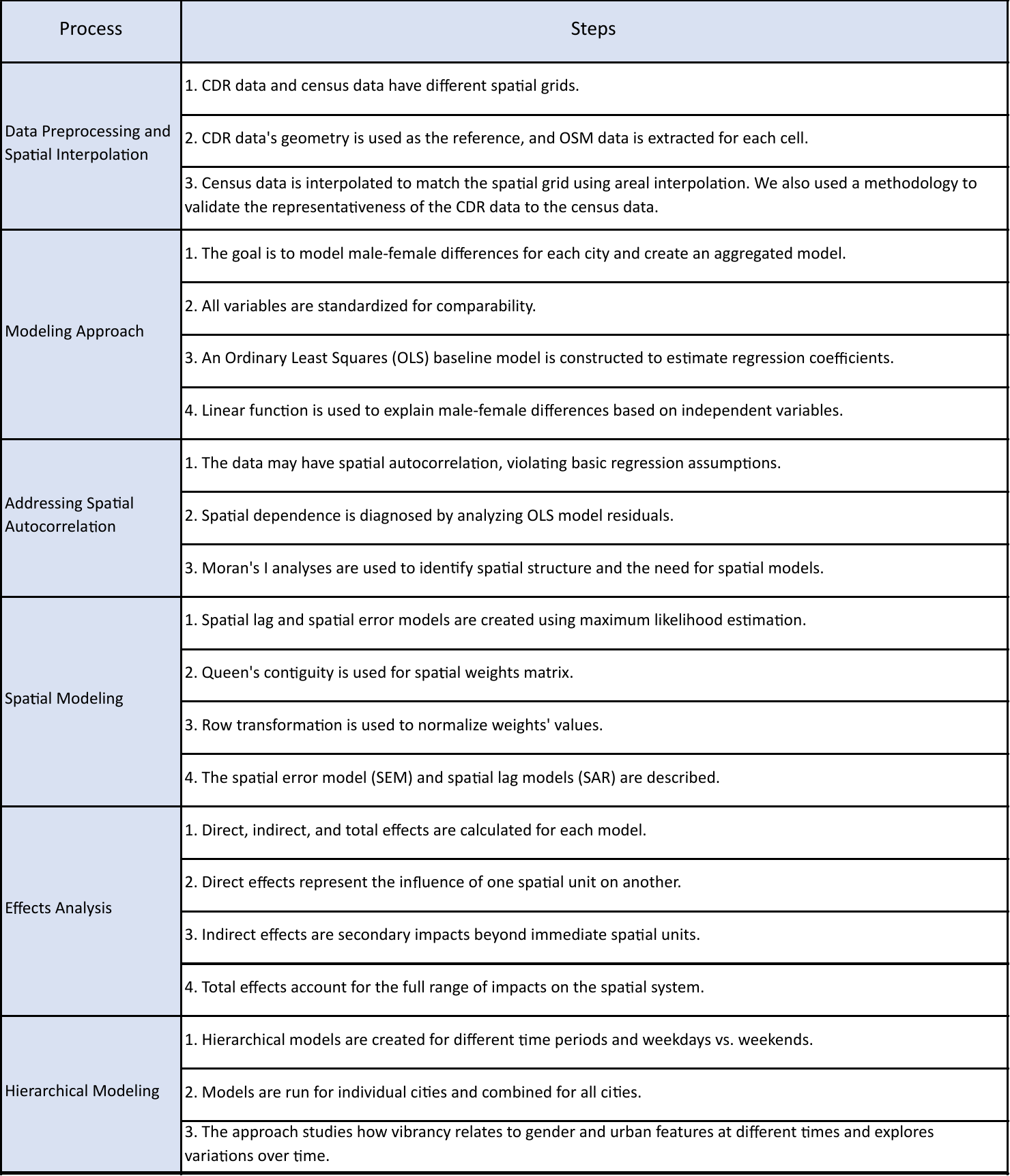}
	\caption{An overview of the steps of the analysis.}
\label{fig:si_1}
\end{figure*}
\begin{figure*}[htp]
	\includegraphics[width=\linewidth]{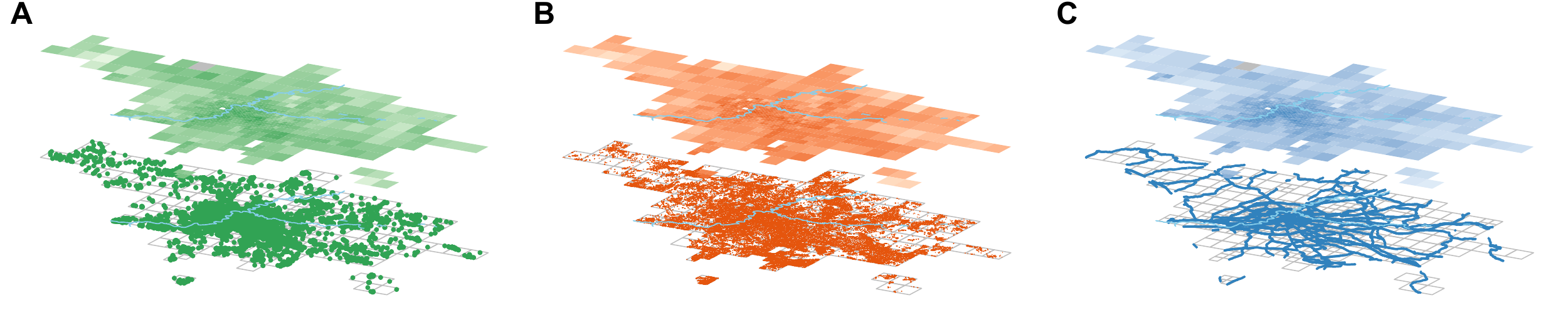}
	\caption{OpenStreetMap (OSM) data types for Roma. OSM provides a range of different data types, from points to lines, and geometries. These features might be associated with male-female differences. In the image, we can see (A) `Points of Interest`, (B) buildings' locations and geometries, and (C) road lengths and intersections. Each of the data types contains detailed information regarding each location. Each of the data types has been used and converted to a density estimate in the above choropleth. OSM data are taken from~\cite{osmOpenStreetMapContributors2017} and are visualised on the Telecom Italia grid for Roma~\citep{gruppotimDatasetSourceTIM2015}.}
\label{fig:si_2}
\end{figure*}
\begin{figure*}[htp]
	\includegraphics[width=\linewidth]{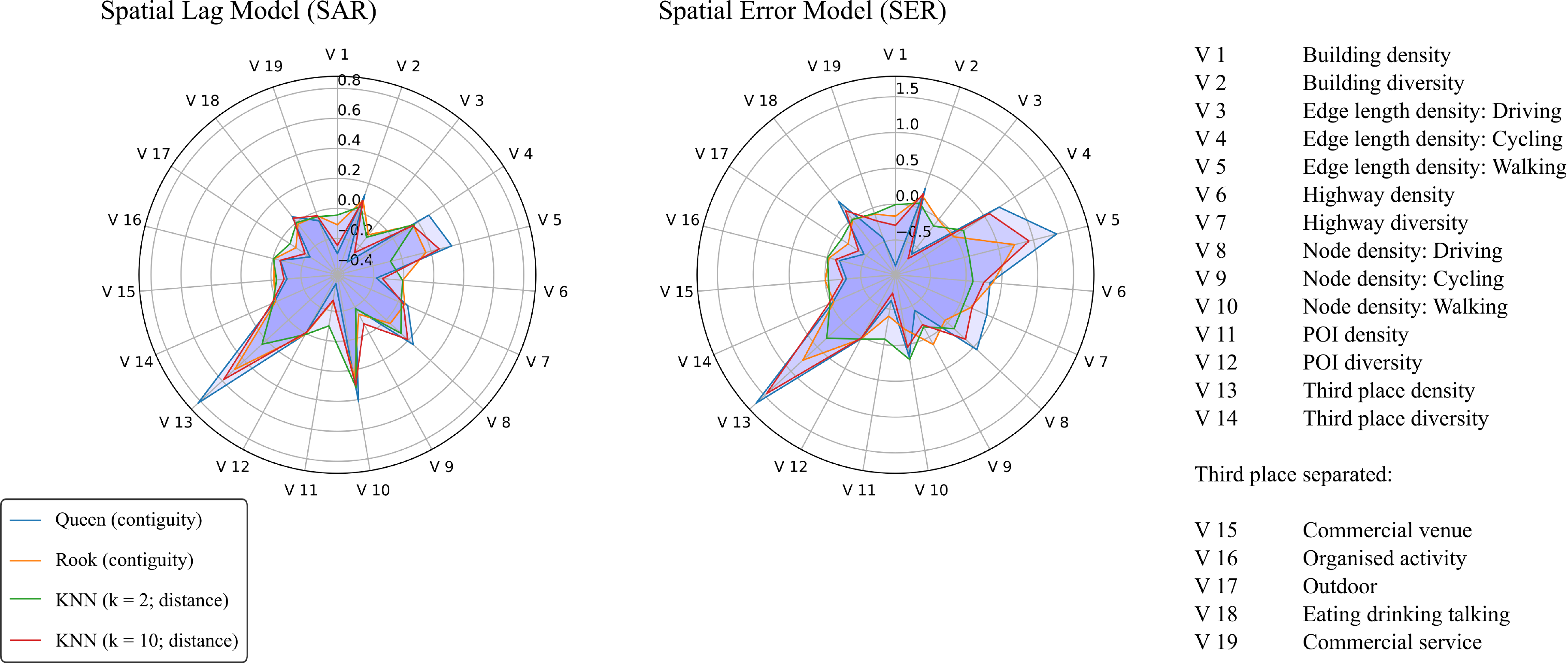}
	\caption{Comparison of the impact of the spatial weights matrix on the indirect effects of the spatial error model and spatial lag models for Roma. The image shows the indirect effects for all variables in the analysis (V 1 -- V 19) when the spatial weights are either Queen's (contiguity), Rook's (contiguity), or k-nearest neighbours (distance) with k values of 2 and 10, respectively.}
\label{fig:si_3}
\end{figure*}
\begin{figure*}[!htb]
	\includegraphics[scale=1.2]{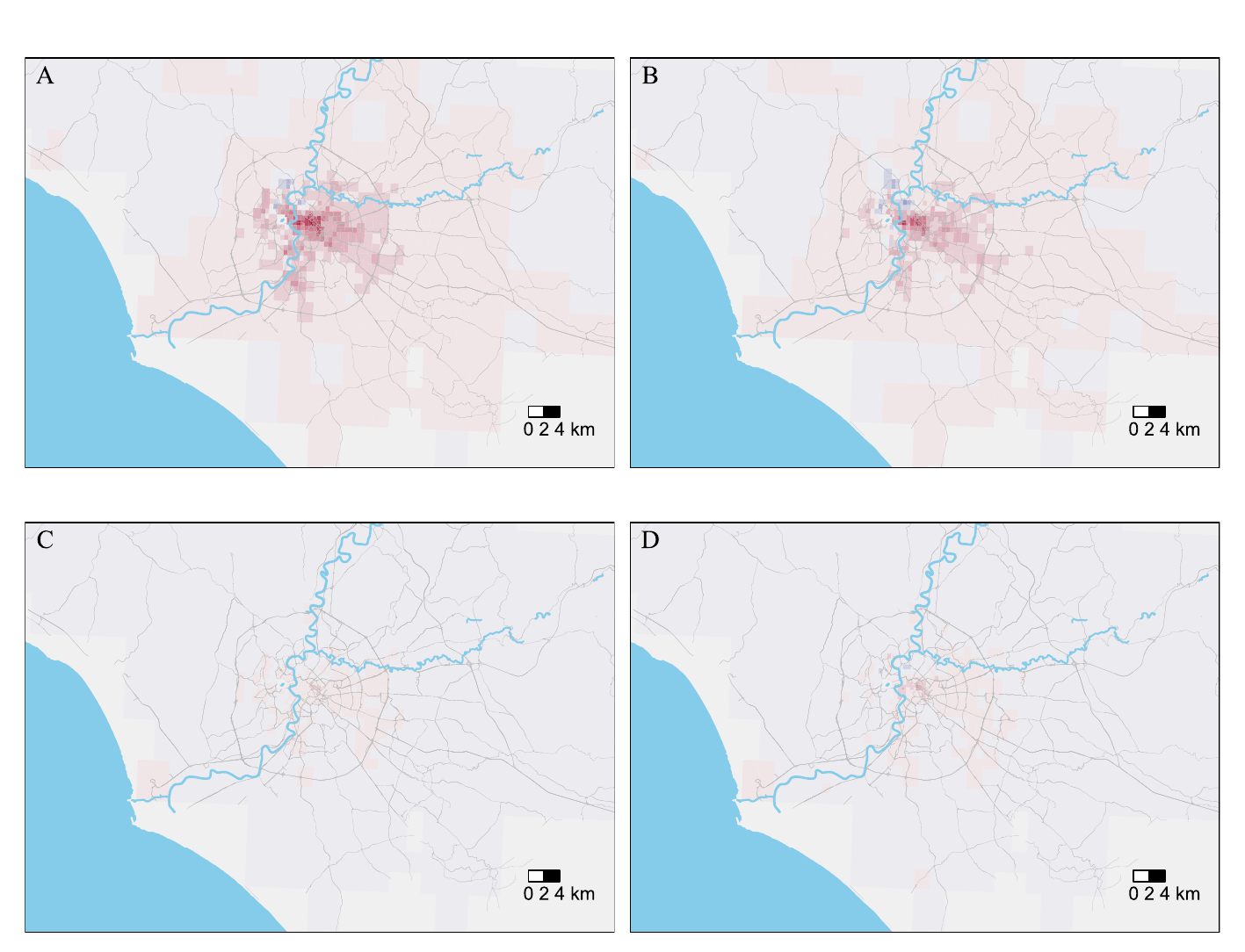}
	\caption[Differences in urban vibrancy across the day]{Roma: Visual inspection suggests the presence of differences between males and females. Each map of Roma shows the male-female differences derived from the Call Detail Record (CDR) data. The data are split into four categories for the periods of (A) Morning (06:00-12:00), (B) Afternoon (12:00-18:00), (C) Evening (18:00-00:00), and (D) Night (00:00-06:00). CDR data taken from ~\cite{gruppotimDatasetSourceTIM2015} and OpenStreetMap data taken from~\cite{osmOpenStreetMapContributors2017}.\hfill \break}\label{fig:si_4}
\end{figure*}
\begin{figure*}[t]
	\includegraphics[width=\linewidth]{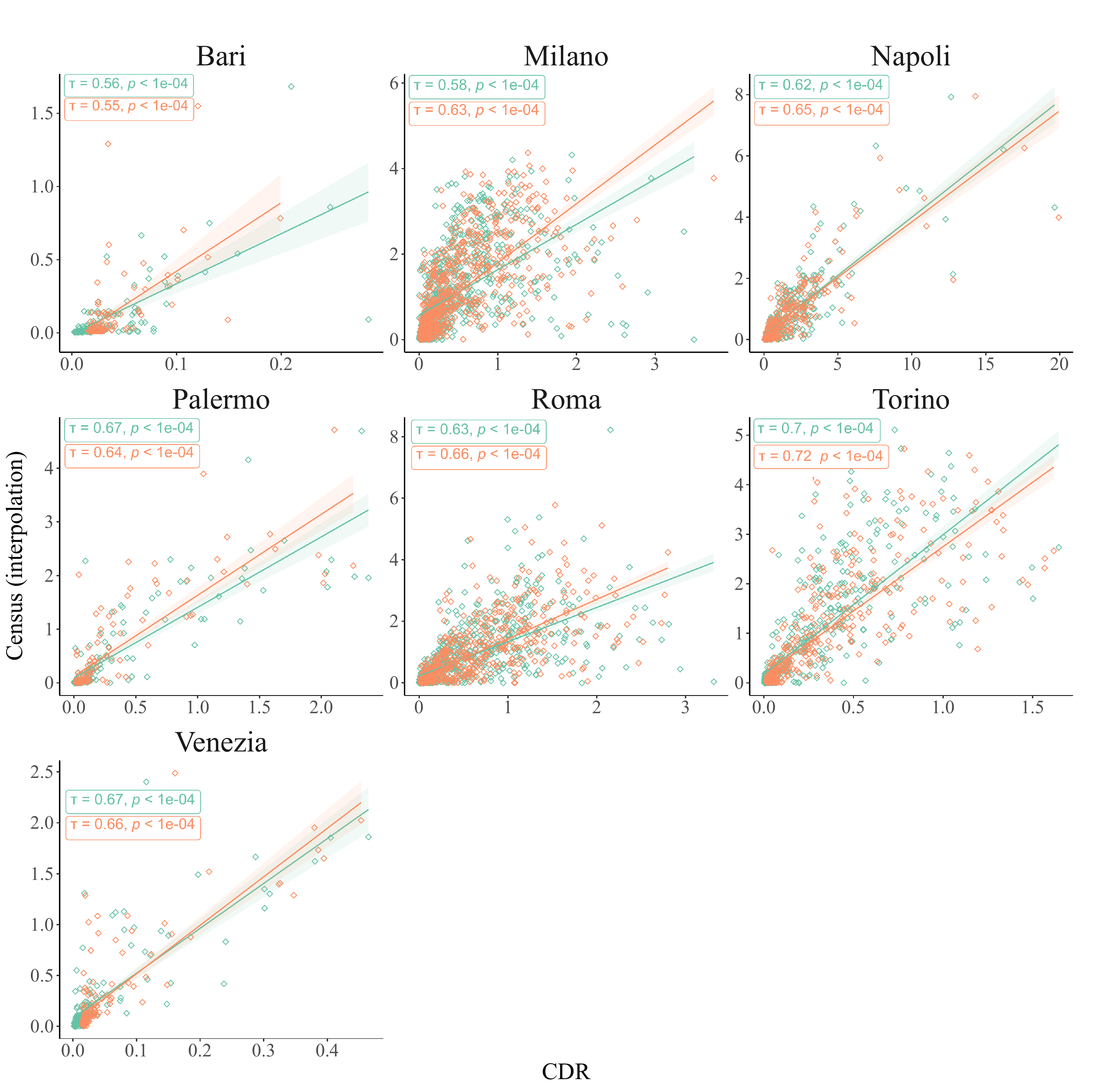}
	\caption{Correlation between Call Detail Record (CDR) data and census data. Orange circles represent females and green circles represent males. We test the correlation of the data sources following the process of areal interpolation using Kendall's tau within a rank correlation analysis~\citep{hamedDistributionKendallTau2011}. This shows a positive relationship between CDR and census data. Cities are presented in alphabetical order.}\label{fig:figure-si_5}
\end{figure*}
\clearpage
\begin{table}[ht]
\caption{Moran's I analysis of spatial dependence for each level of the model hierarchy.}
\centering
\begin{tabular}{lllllllllll}
\toprule
Model hierarchy       & City                       & City    & Moran's I & Moran's I (std) & $p$-value & Sig. star \\ 
\midrule
i   & Day                                          & \texttt{Bari}    & 0.0138    & 0.995           & 0.3197  & ns        \\
                      &                            & \texttt{Milano}  & 0.3826    & 21.0432         & 0       & (****)    \\
                      &                            & \texttt{Napoli}  & 0.0364    & 1.6846          & 0.0921  & (ns)      \\
                      &                            & \texttt{Palermo} & 0.236     & 5.4222          & 0       & (****)    \\
                      &                            & \texttt{Roma}    & 0.3668    & 19.2587         & 0       & (****)    \\
                      &                            & \texttt{Torino}  & 0.2483    & 11.3134         & 0       & (****)    \\
                      &                            & \texttt{Venezia} & 0.1171    & 2.8606          & 0.0042  & (**)      \\
                      &            Night           & \texttt{Bari}    & 0.0312    & 1.3465          & 0.1781  & (ns)      \\
                      &                            & \texttt{Milano}  & 0.2449    & 13.5969         & 0       & (****)    \\
                      &                            & \texttt{Napoli}  & -0.0173   & 0.0219          & 0.9825  & (ns)      \\
                      &                            & \texttt{Palermo} & 0.1501    & 3.6517          & 0.0003  & (***)     \\
                      &                            & \texttt{Roma}    & 0.3463    & 18.2038         & 0       & (****)    \\
                      &                            & \texttt{Torino}  & 0.1526    & 7.1234          & 0       & (****)    \\
                      &                            & \texttt{Venezia} & 0.0931    & 2.3704          & 0.0178  & (*)       \\
ii  & Week                                         & \texttt{Bari}    & 0.0373    & 1.4701          & 0.1415  & (ns)      \\
                      &                            & \texttt{Milano}  & 0.3581    & 19.7161         & 0       & (****)    \\
                      &                            & \texttt{Napoli}  & 0.0418    & 1.8497          & 0.0644  & (ns)      \\
                      &                            & \texttt{Palermo} & 0.2546    & 5.8043          & 0       & (****)    \\
                      &                            & \texttt{Roma}    & 0.3418    & 17.9708         & 0       & (****)    \\
                      &                            & \texttt{Torino}  & 0.2356    & 10.7562         & 0       & (****)    \\
                      &                            & \texttt{Venezia} & 0.1237    & 2.996           & 0.0027  & (**)      \\
                      &  Weekend                   & \texttt{Bari}    & -0.0064   & 0.5879          & 0.5566  & (ns)      \\
                      &                            & \texttt{Milano}  & 0.3852    & 21.182          & 0       & (****)    \\
                      &                            & \texttt{Napoli}  & 0.0135    & 0.9742          & 0.33    & (ns)      \\
                      &                            & \texttt{Palermo} & 0.1932    & 4.5393          & 0       & (****)    \\
                      &                            & \texttt{Roma}    & 0.4012    & 21.0345         & 0       & (****)    \\
                      &                            & \texttt{Torino}  & 0.2436    & 11.1083         & 0       & (****)    \\
                      &                            & \texttt{Venezia} & 0.1055    & 2.6224          & 0.0087  & (**)      \\
iii & Morning                                      & \texttt{Bari}    & 0.002     & 0.7578          & 0.4486  & (ns)      \\
                      &                            & \texttt{Milano}  & 0.3749    & 20.6226         & 0       & (****)    \\
                      &                            & \texttt{Napoli}  & 0.0641    & 2.5403          & 0.0111  & (*)       \\
                      &                            & \texttt{Palermo} & 0.2405    & 5.5141          & 0       & (****)    \\
                      &                            & \texttt{Roma}    & 0.3435    & 18.0568         & 0       & (****)    \\
                      &                            & \texttt{Torino}  & 0.2341    & 10.6906         & 0       & (****)    \\
                      &                            & \texttt{Venezia} & 0.09      & 2.3067          & 0.0211  & (*)       \\
                      &  Afternoon                 & \texttt{Bari}    & 0.034     & 1.4039          & 0.1604  & (ns)      \\
                      &                            & \texttt{Milano}  & 0.3869    & 21.2759         & 0       & (****)    \\
                      &                            & \texttt{Napoli}  & 0.0305    & 1.4994          & 0.1338  & (ns)      \\
                      &                            & \texttt{Palermo} & 0.2368    & 5.4373          & 0       & (****)    \\
                      &                            & \texttt{Roma}    & 0.3852    & 20.2063         & 0       & (****)    \\
                      &                            & \texttt{Torino}  & 0.2581    & 11.7389         & 0       & (****)    \\
                      &                            & \texttt{Venezia} & 0.1489    & 3.511           & 0.0004  & (***)     \\
                      &  Evening                   & \texttt{Bari}    & 0.046     & 1.6445          & 0.1001  & (ns)      \\
                      &                            & \texttt{Milano}  & 0.2519    & 13.9744         & 0       & (****)    \\
                      &                            & \texttt{Napoli}  & -0.0157   & 0.0721          & 0.9425  & (ns)      \\
                      &                            & \texttt{Palermo} & 0.0821    & 2.2483          & 0.0246  & (*)       \\
                      &                            & \texttt{Roma}    & 0.3708    & 19.4634         & 0       & (****)    \\
                      &                            & \texttt{Torino}  & 0.1599    & 7.4434          & 0       & (****)    \\
                      &                            & \texttt{Venezia} & 0.1313    & 3.1517          & 0.0016  & (**)      \\
                      & Night                      & \texttt{Bari}    & 0.0343    & 1.4084          & 0.159   & (ns)      \\
                      &                            & \texttt{Milano}  & 0.2862    & 15.8271         & 0       & (****)    \\
                      &                            & \texttt{Napoli}  & 0.0444    & 1.9297          & 0.0536  & (ns)      \\
                      &                            & \texttt{Palermo} & 0.1683    & 4.0266          & 0.0001  & (****)    \\
                      &                            & \texttt{Roma}    & 0.3702    & 19.433          & 0       & (****)    \\
                      &                            & \texttt{Torino}  & 0.2085    & 9.5716          & 0       & (****)    \\
                      &                            & \texttt{Venezia} & 0.1145    & 2.8076          & 0.005   & (**)      \\ \cmidrule(l){3-7}
\end{tabular}\label{tab:moran-i}
\end{table}
\clearpage
\begin{figure*}[t]
	\includegraphics[width=\linewidth]{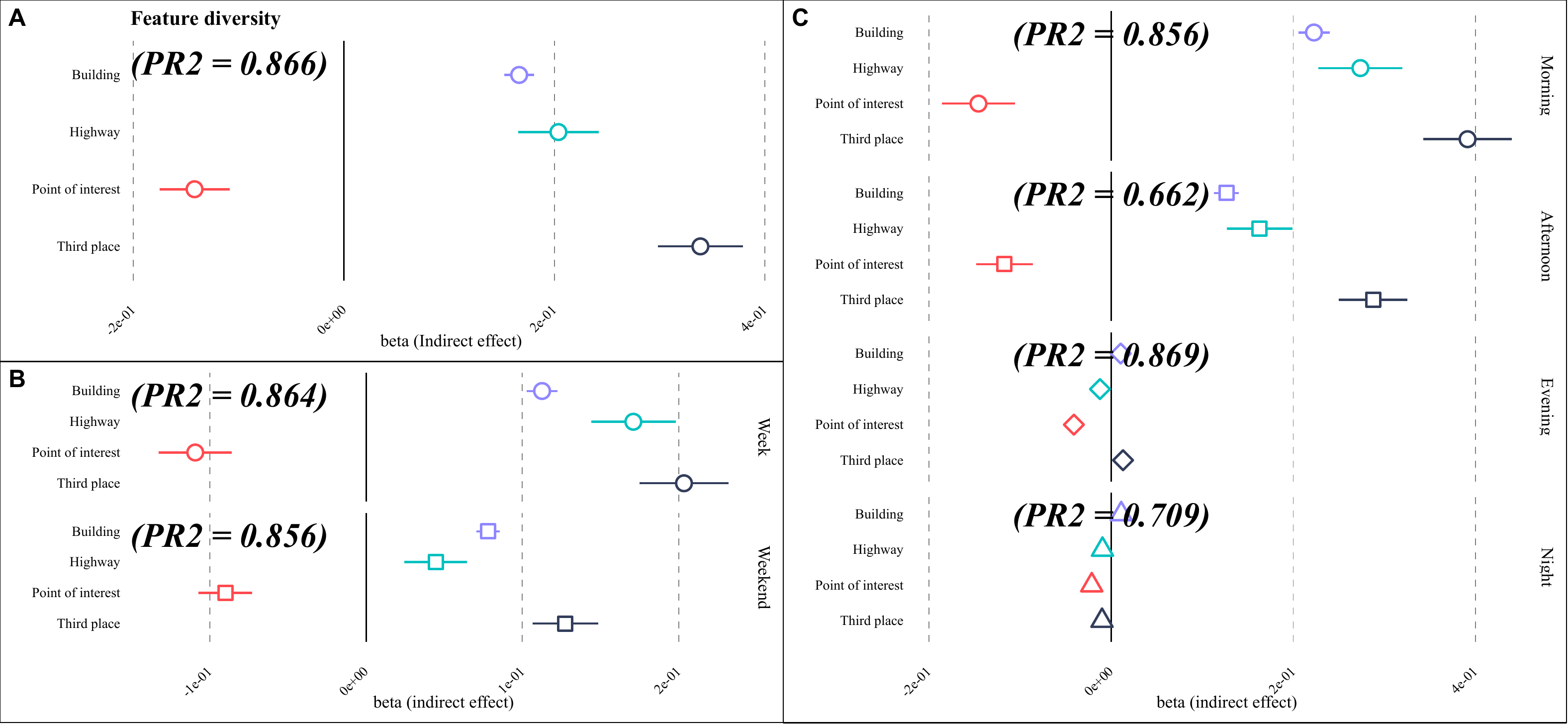}
	\caption{The relationship between diversity in features excluding (`third place` variables--see below) and male-female differences for the spatial model aggregating all cities. The plot shows the significance of the direct effect; however, the \(\beta\) coefficients represent the indirect effect (see Section~\ref{subsec:statistical-approach} for definitions). The plot displays all diversity variables, with the y-axis showing the variables for each model and the colour representing each variable. Panel (A) shows all daytime data between 08:00 and 20:00; Panel (B) displays weekday (Monday-Thursday) versus weekend (Friday-Sunday) data within the same time period; Panel (C) shows data averaged into four categories: morning (06:00-12:00), afternoon (12:00-18:00), evening (18:00-00:00), and night (00:00-06:00). To aid comparison, shapes denote categories in Panels (B) and (C). Significance is indicated by closed and open shapes (\emph{P} \(<\) 0.05 and \emph{P} \(>\) 0.05, respectively), and each shape shows the error bars as a horizontal line. The pseudo-r-squared is reported for each panel.}\label{fig:si_6}
\end{figure*}
\begin{figure*}[t]
	\includegraphics[width=\linewidth]{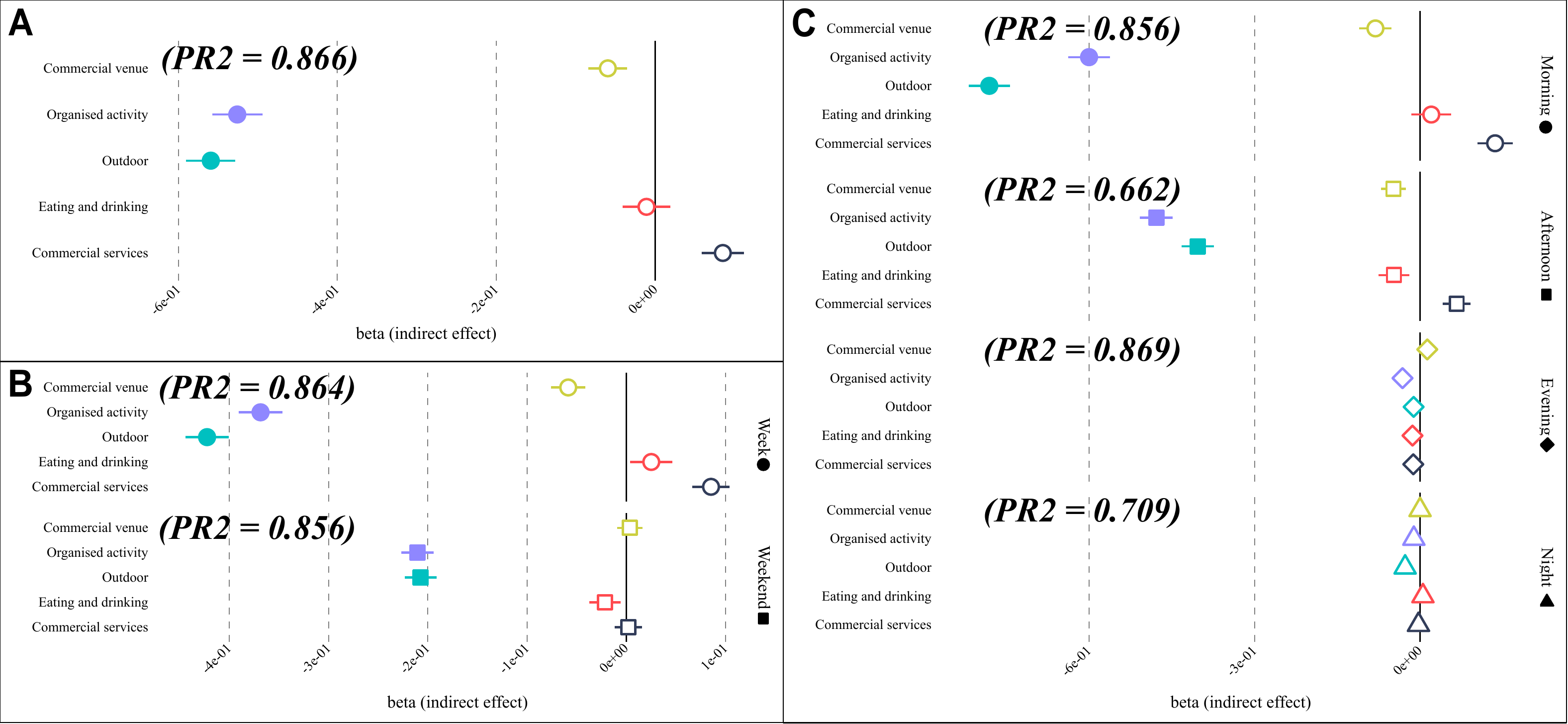}
	\caption{The relationship between diversity in `third place` features and male-female differences for the spatial model aggregating all cities. The plot shows the significance of the direct effect; however, the \(\beta\) coefficients represent the indirect effect (see Section~\ref{subsec:statistical-approach} for definitions). The plot displays all diversity variables, with the y-axis showing the variables for each model and the colour representing each variable. Panel (A) shows all daytime data between 08:00 and 20:00; Panel (B) displays weekday (Monday-Thursday) versus weekend (Friday-Sunday) data within the same time period; Panel (C) shows data averaged into four categories: morning (06:00-12:00), afternoon (12:00-18:00), evening (18:00-00:00), and night (00:00-06:00). To aid comparison, shapes denote categories in Panels (B) and (C). Significance is indicated by closed and open shapes (\emph{P} \(<\) 0.05 and \emph{P} \(>\) 0.05, respectively), and each shape shows the error bars as a horizontal line. The pseudo-r-squared is reported for each panel.}\label{fig:si_7}
\end{figure*}

%% file: 00_main.bbl
\begin{thebibliography}{69}
\providecommand{\natexlab}[1]{#1}
\providecommand{\url}[1]{\texttt{#1}}
\expandafter\ifx\csname urlstyle\endcsname\relax
  \providecommand{\doi}[1]{doi: #1}\else
  \providecommand{\doi}{doi: \begingroup \urlstyle{rm}\Url}\fi

\bibitem[{UN DESA}(2019)]{undesaWorldUrbanizationProspects2019}
{UN DESA}.
\newblock \emph{World {{Urbanization Prospects}}: {{The}} 2018 {{Revision}}}.
\newblock {United Nations}, August 2019.
\newblock ISBN 978-92-1-004314-4.
\newblock \doi{10.18356/b9e995fe-en}.

\bibitem[Harris(1990)]{harrisUrbanisationEconomicDevelopment1990a}
Nigel Harris.
\newblock Urbanisation, economic development and policy in developing
  countries.
\newblock \emph{Habitat International}, 14\penalty0 (4):\penalty0 3--42,
  January 1990.
\newblock ISSN 0197-3975.
\newblock \doi{10.1016/0197-3975(90)90002-I}.

\bibitem[Bettencourt et~al.(2007)Bettencourt, Lobo, Helbing, K{\"u}hnert, and
  West]{bettencourtGrowthInnovationScaling2007a}
Lu{\'i}s M.~A. Bettencourt, Jos{\'e} Lobo, Dirk Helbing, Christian K{\"u}hnert,
  and Geoffrey~B. West.
\newblock Growth, innovation, scaling, and the pace of life in cities.
\newblock \emph{Proceedings of the National Academy of Sciences}, 104\penalty0
  (17):\penalty0 7301--7306, April 2007.
\newblock ISSN 0027-8424, 1091-6490.
\newblock \doi{10.1073/pnas.0610172104}.

\bibitem[Liu et~al.(2020)Liu, Wu, Thakuriah, and
  Wang]{liuGeographyHumanActivity2020}
Wei Liu, Wenjie Wu, Piyushimita Thakuriah, and Jianghao Wang.
\newblock The geography of human activity and land use: {{A}} big data
  approach.
\newblock \emph{Cities}, 97:\penalty0 102523, February 2020.
\newblock ISSN 0264-2751.
\newblock \doi{10.1016/j.cities.2019.102523}.

\bibitem[Thomas(2008)]{thomasUrbanisationDriverChange2008}
S.~Thomas.
\newblock Urbanisation as a driver of change.
\newblock In \emph{{{SUSTAINABLE CITY}} 2008}, pages 95--104, {Skiathos,
  Greece}, August 2008.
\newblock \doi{10.2495/SC080101}.

\bibitem[Nelson et~al.(2002)Nelson, Pendall, Dawkins, and
  Knaap]{nelsonLinkGrowthManagement2002}
A.~Nelson, R.~Pendall, Casey~J. Dawkins, and G.~Knaap.
\newblock The link between growth management and housing affordability: The
  academic evidence.
\newblock Technical report, {The Brookings institution center on urban and
  metropolitan policy}, February 2002.

\bibitem[El~Araby(2002)]{elarabyUrbanGrowthEnvironmental2002}
M~El~Araby.
\newblock Urban growth and environmental degradation.
\newblock \emph{Cities}, 19\penalty0 (6):\penalty0 389--400, December 2002.
\newblock ISSN 02642751.
\newblock \doi{10.1016/S0264-2751(02)00069-0}.

\bibitem[Glaeser and Sacerdote(1999)]{glaeserWhyThereMore1999a}
Edward~L. Glaeser and Bruce Sacerdote.
\newblock Why is {{There More Crime}} in {{Cities}}?
\newblock \emph{Journal of Political Economy}, 107\penalty0 (S6):\penalty0
  S225--S258, 1999.
\newblock ISSN 0022-3808.
\newblock \doi{10.1086/250109}.

\bibitem[Connolly et~al.(2021)Connolly, Keil, and
  Ali]{connollyExtendedUrbanisationSpatialities2021}
Creighton Connolly, Roger Keil, and S.~Harris Ali.
\newblock Extended urbanisation and the spatialities of infectious disease:
  {{Demographic}} change, infrastructure and governance.
\newblock \emph{Urban Studies}, 58\penalty0 (2):\penalty0 245--263, February
  2021.
\newblock ISSN 0042-0980, 1360-063X.
\newblock \doi{10.1177/0042098020910873}.

\bibitem[Santana et~al.(2023)Santana, Botta, Barbosa, Privitera, Menezes, and
  {Di Clemente}]{Santana2022}
Clodomir Santana, Federico Botta, Hugo Barbosa, Filippo Privitera, Ronaldo
  Menezes, and Riccardo {Di Clemente}.
\newblock {COVID-19 is linked to changes in the time–space dimension of human
  mobility}.
\newblock \emph{Nature Human Behaviour}, jul 2023.
\newblock ISSN 2397-3374.
\newblock \doi{10.1038/s41562-023-01660-3}.

\bibitem[Zheng et~al.(2014)Zheng, Capra, Wolfson, and
  Yang]{zhengUrbanComputingConcepts2014a}
Yu~Zheng, Licia Capra, Ouri Wolfson, and Hai Yang.
\newblock Urban {{Computing}}: {{Concepts}}, {{Methodologies}}, and
  {{Applications}}.
\newblock \emph{ACM Transactions on Intelligent Systems and Technology},
  5\penalty0 (3):\penalty0 1--55, October 2014.
\newblock ISSN 2157-6904, 2157-6912.
\newblock \doi{10.1145/2629592}.

\bibitem[Bassolas et~al.(2019)Bassolas, {Barbosa-Filho}, Dickinson, Dotiwalla,
  Eastham, Gallotti, Ghoshal, Gipson, Hazarie, Kautz, Kucuktunc, Lieber,
  Sadilek, and Ramasco]{bassolasHierarchicalOrganizationUrban2019a}
Aleix Bassolas, Hugo {Barbosa-Filho}, Brian Dickinson, Xerxes Dotiwalla, Paul
  Eastham, Riccardo Gallotti, Gourab Ghoshal, Bryant Gipson, Surendra~A.
  Hazarie, Henry Kautz, Onur Kucuktunc, Allison Lieber, Adam Sadilek, and
  Jos{\'e}~J. Ramasco.
\newblock Hierarchical organization of urban mobility and its connection with
  city livability.
\newblock \emph{Nature Communications}, 10\penalty0 (1):\penalty0 4817, October
  2019.
\newblock ISSN 2041-1723.
\newblock \doi{10.1038/s41467-019-12809-y}.

\bibitem[Kalila et~al.(2018)Kalila, Awwad, {Di Clemente}, and
  Gonz{\'{a}}lez]{Kalila2018}
Adham Kalila, Zeyad Awwad, Riccardo {Di Clemente}, and Marta~C. Gonz{\'{a}}lez.
\newblock {Big data fusion to estimate urban fuel consumption: A case study of
  riyadh}.
\newblock \emph{Transportation Research Record}, 2672\penalty0 (24):\penalty0
  49--59, dec 2018.
\newblock ISSN 21694052.
\newblock \doi{10.1177/0361198118798461}.

\bibitem[Xu et~al.(2021)Xu, {Di Clemente}, and Gonz{\'{a}}lez]{Xu2021}
Yanyan Xu, Riccardo {Di Clemente}, and Marta~C. Gonz{\'{a}}lez.
\newblock {Understanding vehicular routing behavior with location-based service
  data}.
\newblock \emph{EPJ Data Science}, 10\penalty0 (1):\penalty0 12, dec 2021.
\newblock ISSN 2193-1127.
\newblock \doi{10.1140/epjds/s13688-021-00267-w}.

\bibitem[T{\'o}th et~al.(2021)T{\'o}th, Wachs, Di~Clemente, Jakobi,
  S{\'a}gv{\'a}ri, Kert{\'e}sz, and Lengyel]{tothInequalityRisingWhere2021}
Gerg{\H o} T{\'o}th, Johannes Wachs, Riccardo Di~Clemente, {\'A}kos Jakobi,
  Bence S{\'a}gv{\'a}ri, J{\'a}nos Kert{\'e}sz, and Bal{\'a}zs Lengyel.
\newblock Inequality is rising where social network segregation interacts with
  urban topology.
\newblock \emph{Nature Communications}, 12\penalty0 (1):\penalty0 1143,
  February 2021.
\newblock ISSN 2041-1723.
\newblock \doi{10.1038/s41467-021-21465-0}.

\bibitem[Bor et~al.(2017)Bor, Cohen, and Galea]{borPopulationHealthEra2017}
Jacob Bor, Gregory~H Cohen, and Sandro Galea.
\newblock Population health in an era of rising income inequality: {{USA}},
  1980\textendash 2015.
\newblock \emph{The Lancet}, 389\penalty0 (10077):\penalty0 1475--1490, April
  2017.
\newblock ISSN 01406736.
\newblock \doi{10.1016/S0140-6736(17)30571-8}.

\bibitem[Weller and {van
  Hulten}(2012)]{wellerGentrificationDisplacementEffects2012}
Sally Weller and Andrew {van Hulten}.
\newblock Gentrification and {{Displacement}}: {{The Effects}} of a {{Housing
  Crisis}} on {{Melbourne}}'s {{Low-Income Residents}}.
\newblock \emph{Urban Policy and Research}, 30\penalty0 (1):\penalty0 25--42,
  March 2012.
\newblock ISSN 0811-1146.
\newblock \doi{10.1080/08111146.2011.635410}.

\bibitem[Entwisle(2007)]{entwislePuttingPeoplePlace2007b}
Barbara Entwisle.
\newblock Putting people into place.
\newblock \emph{Demography}, 44\penalty0 (4):\penalty0 687--703, November 2007.
\newblock ISSN 0070-3370.
\newblock \doi{10.1353/dem.2007.0045}.

\bibitem[King(2013)]{kingJaneJacobsNeed2013a}
Katherine King.
\newblock Jane {{Jacobs}} and `{{The Need}} for {{Aged Buildings}}':
  {{Neighbourhood Historical Development Pace}} and {{Community Social
  Relations}}.
\newblock \emph{Urban Studies}, 50\penalty0 (12):\penalty0 2407--2424,
  September 2013.
\newblock ISSN 0042-0980.
\newblock \doi{10.1177/0042098013477698}.

\bibitem[Steenbruggen et~al.(2015)Steenbruggen, Tranos, and
  Nijkamp]{steenbruggenDataMobilePhone2015}
John Steenbruggen, Emmanouil Tranos, and Peter Nijkamp.
\newblock Data from mobile phone operators: {{A}} tool for smarter cities?
\newblock \emph{Telecommunications Policy}, 39\penalty0 (3):\penalty0 335--346,
  May 2015.
\newblock ISSN 0308-5961.
\newblock \doi{10.1016/j.telpol.2014.04.001}.

\bibitem[Gonz{\'a}lez et~al.(2008)Gonz{\'a}lez, Hidalgo, and
  Barab{\'a}si]{gonzalezUnderstandingIndividualHuman2008}
Marta~C. Gonz{\'a}lez, C{\'e}sar~A. Hidalgo, and Albert-L{\'a}szl{\'o}
  Barab{\'a}si.
\newblock Understanding individual human mobility patterns.
\newblock \emph{Nature}, 453\penalty0 (7196):\penalty0 779--782, June 2008.
\newblock ISSN 1476-4687.
\newblock \doi{10.1038/nature06958}.

\bibitem[Lazer et~al.(2009)Lazer, Pentland, Adamic, Aral, Barabasi, Brewer,
  Christakis, Contractor, Fowler, Gutmann, Jebara, King, Macy, Roy, and
  Van~Alstyne]{lazerLifeNetworkComing2009a}
David Lazer, Alex~(Sandy) Pentland, Lada Adamic, Sinan Aral, Albert~Laszlo
  Barabasi, Devon Brewer, Nicholas Christakis, Noshir Contractor, James Fowler,
  Myron Gutmann, Tony Jebara, Gary King, Michael Macy, Deb Roy, and Marshall
  Van~Alstyne.
\newblock Life in the network: The coming age of computational social science.
\newblock \emph{Science (New York, N.Y.)}, 323\penalty0 (5915):\penalty0
  721--723, February 2009.
\newblock ISSN 0036-8075.
\newblock \doi{10.1126/science.1167742}.

\bibitem[Vespignani(2009)]{vespignaniPredictingBehaviorTechnoSocial2009}
Alessandro Vespignani.
\newblock Predicting the {{Behavior}} of {{Techno-Social Systems}}.
\newblock \emph{Science}, 325\penalty0 (5939):\penalty0 425--428, July 2009.
\newblock \doi{10.1126/science.1171990}.

\bibitem[Salesses et~al.(2013)Salesses, Schechtner, and
  Hidalgo]{salessesCollaborativeImageCity2013a}
Philip Salesses, Katja Schechtner, and C{\'e}sar~A. Hidalgo.
\newblock The {{Collaborative Image}} of {{The City}}: {{Mapping}} the
  {{Inequality}} of {{Urban Perception}}.
\newblock \emph{PLOS ONE}, 8\penalty0 (7):\penalty0 e68400, July 2013.
\newblock ISSN 1932-6203.
\newblock \doi{10.1371/journal.pone.0068400}.

\bibitem[Botta et~al.(2015)Botta, Moat, and
  Preis]{bottaQuantifyingCrowdSize2015b}
Federico Botta, Helen~Susannah Moat, and Tobias Preis.
\newblock Quantifying crowd size with mobile phone and {{{\emph{Twitter}}}}
  data.
\newblock \emph{Royal Society Open Science}, 2\penalty0 (5):\penalty0 150162,
  May 2015.
\newblock ISSN 2054-5703.
\newblock \doi{10.1098/rsos.150162}.

\bibitem[Seresinhe et~al.(2016)Seresinhe, Preis, and
  Moat]{seresinheQuantifyingLinkArt2016}
Chanuki~Illushka Seresinhe, Tobias Preis, and Helen~Susannah Moat.
\newblock Quantifying the link between art and property prices in urban
  neighbourhoods.
\newblock \emph{Royal Society Open Science}, 3\penalty0 (4):\penalty0 160146,
  April 2016.
\newblock \doi{10.1098/rsos.160146}.

\bibitem[Preis et~al.(2020)Preis, Botta, and
  Moat]{preisSensingGlobalTourism2020}
Tobias Preis, Federico Botta, and Helen~Susannah Moat.
\newblock Sensing global tourism numbers with millions of publicly shared
  online photographs.
\newblock \emph{Environment and Planning A: Economy and Space}, 52\penalty0
  (3):\penalty0 471--477, May 2020.
\newblock ISSN 0308-518X.
\newblock \doi{10.1177/0308518X19872772}.

\bibitem[Di~Clemente et~al.(2018)Di~Clemente, {Luengo-Oroz}, Travizano, Xu,
  Vaitla, and Gonz{\'a}lez]{diclementeSequencesPurchasesCredit2018a}
Riccardo Di~Clemente, Miguel {Luengo-Oroz}, Matias Travizano, Sharon Xu, Bapu
  Vaitla, and Marta~C. Gonz{\'a}lez.
\newblock Sequences of purchases in credit card data reveal lifestyles in urban
  populations.
\newblock \emph{Nature Communications}, 9\penalty0 (1):\penalty0 3330, August
  2018.
\newblock ISSN 2041-1723.
\newblock \doi{10.1038/s41467-018-05690-8}.

\bibitem[Bannister and Botta(2021)]{bannisterRapidIndicatorsDeprivation2021a}
Adam Bannister and Federico Botta.
\newblock Rapid indicators of deprivation using grocery shopping data.
\newblock \emph{Royal Society Open Science}, 8\penalty0 (12):\penalty0 211069,
  December 2021.
\newblock \doi{10.1098/rsos.211069}.

\bibitem[Xu et~al.(2019)Xu, {Di Clemente}, and Gonz{\'{a}}lez]{Xu2019}
Sharon Xu, Riccardo {Di Clemente}, and Marta~C. Gonz{\'{a}}lez.
\newblock {Mining urban lifestyles: Urban computing, human behavior and
  recommender systems}.
\newblock In \emph{Big Data Recommender Systems: Application Paradigms}, pages
  71--81. Institution of Engineering and Technology, jul 2019.
\newblock ISBN 9781785619779.
\newblock \doi{10.1049/PBPC035G_ch5}.

\bibitem[Su et~al.(2022)Su, Sun, Fan, Noyman, Pentland, and
  Moro]{suRhythmStreetsStreet2022}
Tianyu Su, Maoran Sun, Zhuangyuan Fan, Ariel Noyman, Alex Pentland, and Esteban
  Moro.
\newblock Rhythm of the streets: A street classification framework based on
  street activity patterns.
\newblock \emph{EPJ Data Science}, 11\penalty0 (1):\penalty0 43, December 2022.
\newblock ISSN 2193-1127.
\newblock \doi{10.1140/epjds/s13688-022-00355-5}.

\bibitem[Batty(2013)]{battyNewScienceCities2013a}
Michael Batty.
\newblock \emph{The {{New Science}} of {{Cities}}}.
\newblock {The MIT Press}, 2013.
\newblock ISBN 978-0-262-01952-1.

\bibitem[Pan et~al.(2013)Pan, Ghoshal, Krumme, Cebrian, and
  Pentland]{panUrbanCharacteristicsAttributable2013a}
Wei Pan, Gourab Ghoshal, Coco Krumme, Manuel Cebrian, and Alex Pentland.
\newblock Urban characteristics attributable to density-driven tie formation.
\newblock \emph{Nature Communications}, 4\penalty0 (1):\penalty0 1961, June
  2013.
\newblock ISSN 2041-1723.
\newblock \doi{10.1038/ncomms2961}.

\bibitem[Barthelemy(2016)]{barthelemyStructureDynamicsCities2016a}
Marc Barthelemy.
\newblock \emph{The {{Structure}} and {{Dynamics}} of {{Cities}}: {{Urban Data
  Analysis}} and {{Theoretical Modeling}}}.
\newblock {Cambridge University Press}, November 2016.
\newblock ISBN 978-1-107-10917-9.
\newblock \doi{10.1017/9781316271377}.

\bibitem[Botta et~al.(2020)Botta, Preis, and Moat]{bottaSearchArtRapid2020a}
Federico Botta, Tobias Preis, and Helen~Susannah Moat.
\newblock In search of art: Rapid estimates of gallery and museum visits using
  {{Google Trends}}.
\newblock \emph{EPJ Data Science}, 9\penalty0 (1):\penalty0 1--12, December
  2020.
\newblock ISSN 2193-1127.
\newblock \doi{10.1140/epjds/s13688-020-00232-z}.

\bibitem[Sulis et~al.(2018)Sulis, Manley, Zhong, and
  Batty]{sulisUsingMobilityData2018}
Patrizia Sulis, Ed~Manley, Chen Zhong, and Michael Batty.
\newblock Using mobility data as proxy for measuring urban vitality.
\newblock \emph{Journal of Spatial Information Science}, \penalty0
  (16):\penalty0 137--162, June 2018.
\newblock ISSN 1948-660X.

\bibitem[Botta and {Guti{\'e}rrez-Roig}(2021)]{bottaModellingUrbanVibrancy2021}
Federico Botta and Mario {Guti{\'e}rrez-Roig}.
\newblock Modelling urban vibrancy with mobile phone and {{OpenStreetMap}}
  data.
\newblock \emph{PLoS ONE}, 16\penalty0 (6 June):\penalty0 1--19, 2021.
\newblock ISSN 19326203.
\newblock \doi{10.1371/journal.pone.0252015}.

\bibitem[Wang et~al.(2021)Wang, Liu, Wang, and
  Fu]{wangMeasuringUrbanVibrancy2021}
Pengyang Wang, Kunpeng Liu, Dongjie Wang, and Yanjie Fu.
\newblock Measuring {{Urban Vibrancy}} of {{Residential Communities Using Big
  Crowdsourced Geotagged Data}}.
\newblock \emph{Frontiers in Big Data}, 4, 2021.
\newblock ISSN 2624-909X.

\bibitem[Jacobs(1961)]{jacobsDeathLifeGreat1961}
Jane Jacobs.
\newblock \emph{The {{Death}} and {{Life}} of {{Great American Cities}}}.
\newblock {Random House;}, 1961.

\bibitem[Moroni(2016)]{moroniUrbanDensityJane2016}
Stefano Moroni.
\newblock Urban density after {{Jane Jacobs}}: The crucial role of diversity
  and emergence.
\newblock \emph{City, Territory and Architecture}, 3\penalty0 (1):\penalty0 13,
  September 2016.
\newblock ISSN 2195-2701.
\newblock \doi{10.1186/s40410-016-0041-1}.

\bibitem[Perrone(2019)]{perroneDowntownPeopleStreetlevel2019}
Camilla Perrone.
\newblock `{{Downtown Is}} for {{People}}': {{The}} street-level approach in
  {{Jane Jacobs}}' legacy and its resonance in the planning debate within the
  complexity theory of cities.
\newblock \emph{Cities}, 91:\penalty0 10--16, August 2019.
\newblock ISSN 0264-2751.
\newblock \doi{10.1016/j.cities.2018.12.023}.

\bibitem[Vaitla et~al.(2017)Vaitla, Bosco, Alegana, Bird, Pezzulo, Hornby,
  Sorichetta, Steele, Ruktanonchai, Ruktanonchai, Wetter, Bengtsson, Tatem, {Di
  Clemente}, Luengo-Oroz, and Gonz{\'{a}}lez]{vaitla_data_good}
Bapu Vaitla, Claudio Bosco, Victor Alegana, Tom Bird, Carla Pezzulo, Graeme
  Hornby, Alessandro Sorichetta, Jessica Steele, Cori Ruktanonchai, Nick
  Ruktanonchai, Erik Wetter, Linus Bengtsson, Andrew~J Tatem, Riccardo {Di
  Clemente}, Miguel Luengo-Oroz, and Marta~C Gonz{\'{a}}lez.
\newblock {Big Data and the Well-Being of Women and Girls Applications on the
  Social Scientific Frontier}.
\newblock \emph{Data2x}, 2017.

\bibitem[{OSM}(2017)]{osmOpenStreetMapContributors2017}
{OSM}.
\newblock {{OpenStreetMap}} contributors.
\newblock https://www.openstreetmap.org/, 2017.

\bibitem[{ISTAT}(2023)]{istatBordersAdministrativeUnits2023}
{ISTAT}.
\newblock Borders of administrative units for statistical purposes as of 1
  {{January}} 2023.
\newblock https://www.istat.it/it/archivio/222527, February 2023.

\bibitem[{GruppoTIM}(2015)]{gruppotimDatasetSourceTIM2015}
{GruppoTIM}.
\newblock Dataset {{Source}}: {{TIM Big Data Challenge}},
  {{https://www.gruppotim.it/en.html.}}
\newblock http://localhost:4000/cases/telecom-italias-big-data-challenge.html,
  2015.

\bibitem[Wang et~al.(2018)Wang, Fu, Zhang, Li, and
  Lin]{wangLearningUrbanCommunity2018a}
Pengyang Wang, Yanjie Fu, Jiawei Zhang, Xiaolin Li, and Dan Lin.
\newblock Learning urban community structures: A collective embedding
  perspective with periodic spatial-temporal mobility graphs.
\newblock \emph{{ACM} Transactions on Intelligent Systems and Technology},
  9\penalty0 (6):\penalty0 63:1--63:28, 2018.
\newblock ISSN 2157-6904.
\newblock \doi{10.1145/3209686}.
\newblock URL \url{https://dl.acm.org/doi/10.1145/3209686}.

\bibitem[Sung et~al.(2013)Sung, Go, and Choi]{sungEvidenceJacobsStreet2013}
Hyun-Gun Sung, Doo-Hwan Go, and Chang~Gyu Choi.
\newblock Evidence of {{Jacobs}}'s street life in the great {{Seoul}} city:
  {{Identifying}} the association of physical environment with walking activity
  on streets.
\newblock \emph{Cities}, 35:\penalty0 164--173, December 2013.
\newblock ISSN 0264-2751.
\newblock \doi{10.1016/j.cities.2013.07.010}.

\bibitem[Yu et~al.(2022)Yu, Cui, Liu, Luo, Tian, and
  Yang]{yuIntergenerationalDifferencesUrban2022a}
Bingjie Yu, Xu~Cui, Runze Liu, Pinyang Luo, Fangzhou Tian, and Tian Yang.
\newblock Intergenerational differences in the urban vibrancy of {{TOD}}:
  {{Impacts}} of the built environment on the activities of different age
  groups.
\newblock \emph{Frontiers in Public Health}, 10, 2022.
\newblock ISSN 2296-2565.

\bibitem[Chen et~al.(2022)Chen, Zhao, Xiao, and
  Lu]{chenInvestigatingSpatiotemporalPattern2022a}
Long Chen, Lingyu Zhao, Yang Xiao, and Yi~Lu.
\newblock Investigating the spatiotemporal pattern between the built
  environment and urban vibrancy using big data in {{Shenzhen}}, {{China}}.
\newblock \emph{Computers, Environment and Urban Systems}, 95:\penalty0 101827,
  July 2022.
\newblock ISSN 0198-9715.
\newblock \doi{10.1016/j.compenvurbsys.2022.101827}.

\bibitem[Shannon(1948)]{shannonMathematicalTheoryCommunication1948}
C.~E. Shannon.
\newblock A mathematical theory of communication.
\newblock \emph{The Bell System Technical Journal}, 27\penalty0 (3):\penalty0
  379--423, July 1948.
\newblock ISSN 0005-8580.
\newblock \doi{10.1002/j.1538-7305.1948.tb01338.x}.

\bibitem[Moro et~al.(2021)Moro, Calacci, Dong, and
  Pentland]{moroMobilityPatternsAre2021b}
Esteban Moro, Dan Calacci, Xiaowen Dong, and Alex Pentland.
\newblock Mobility patterns are associated with experienced income segregation
  in large {{US}} cities.
\newblock \emph{Nature Communications}, 12\penalty0 (1):\penalty0 4633, July
  2021.
\newblock ISSN 2041-1723.
\newblock \doi{10.1038/s41467-021-24899-8}.

\bibitem[Fan et~al.(2022)Fan, Su, Sun, Noyman, Zhang, Pentland, and
  Moro]{fanDiversityDensityExperienced2022}
Zhuangyuan Fan, Tianyu Su, Maoran Sun, Ariel Noyman, Fan Zhang, Alex~Sandy
  Pentland, and Esteban Moro.
\newblock Diversity beyond density: Experienced social mixing of urban streets,
  September 2022.

\bibitem[Oldenburg and Brissett(1982)]{oldenburgThirdPlace1982b}
Ramon Oldenburg and Dennis Brissett.
\newblock The third place.
\newblock \emph{Qualitative Sociology}, 5\penalty0 (4):\penalty0 265--284,
  December 1982.
\newblock ISSN 1573-7837.
\newblock \doi{10.1007/BF00986754}.

\bibitem[Jeffres et~al.(2009)Jeffres, Bracken, Jian, and
  Casey]{jeffresImpactThirdPlaces2009a}
Leo~W. Jeffres, Cheryl~C. Bracken, Guowei Jian, and Mary~F. Casey.
\newblock The {{Impact}} of {{Third Places}} on {{Community Quality}} of
  {{Life}}.
\newblock \emph{Applied Research in Quality of Life}, 4\penalty0 (4):\penalty0
  333--345, December 2009.
\newblock ISSN 1871-2584, 1871-2576.
\newblock \doi{10.1007/s11482-009-9084-8}.

\bibitem[Comber and Zeng(2019)]{comberSpatialInterpolationUsing2019a}
Alexis Comber and Wen Zeng.
\newblock Spatial interpolation using areal features: {{A}} review of methods
  and opportunities using new forms of data with coded illustrations.
\newblock \emph{Geography Compass}, 13\penalty0 (10):\penalty0 e12465, 2019.
\newblock ISSN 1749-8198.
\newblock \doi{10.1111/gec3.12465}.

\bibitem[Bergroth et~al.(2022)Bergroth, J{\"a}rv, Tenkanen, Manninen, and
  Toivonen]{bergroth24hourPopulationDistribution2022a}
Claudia Bergroth, Olle J{\"a}rv, Henrikki Tenkanen, Matti Manninen, and Tuuli
  Toivonen.
\newblock A 24-hour population distribution dataset based on mobile phone data
  from {{Helsinki Metropolitan Area}}, {{Finland}}.
\newblock \emph{Scientific Data}, 9\penalty0 (1):\penalty0 39, February 2022.
\newblock ISSN 2052-4463.
\newblock \doi{10.1038/s41597-021-01113-4}.

\bibitem[Hamed(2011)]{hamedDistributionKendallTau2011}
K.~H. Hamed.
\newblock The distribution of {{Kendall}}'s tau for testing the significance of
  cross-correlation in persistent data.
\newblock \emph{Hydrological Sciences Journal}, 56\penalty0 (5):\penalty0
  841--853, July 2011.
\newblock ISSN 0262-6667, 2150-3435.
\newblock \doi{10.1080/02626667.2011.586948}.

\bibitem[Ward and Gleditsch(2019)]{wardSpatialRegressionModels2019}
Michael~Don Ward and Kristian~Skrede Gleditsch.
\newblock \emph{Spatial Regression Models}.
\newblock Quantitative Applications in the Social Sciences. {SAGE
  Publications}, {Thousand Oaks, California}, second edition edition, 2019.
\newblock ISBN 978-1-5443-2883-6.

\bibitem[Anselin et~al.(2006)Anselin, Syabri, and
  Kho]{anselinGeoDaIntroductionSpatial2006}
Luc Anselin, Ibnu Syabri, and Youngihn Kho.
\newblock {{GeoDa}} : {{An Introduction}} to {{Spatial Data Analysis}}.
\newblock \emph{Geographical Analysis}, 38\penalty0 (1):\penalty0 5--22, 2006.
\newblock ISSN 1538-4632.
\newblock \doi{10.1111/j.0016-7363.2005.00671.x}.

\bibitem[Rey and Anselin(2010)]{reyPySALPythonLibrary2010}
Sergio~J. Rey and Luc Anselin.
\newblock {{PySAL}}: {{A Python Library}} of {{Spatial Analytical Methods}}.
\newblock In Manfred~M. Fischer and Arthur Getis, editors, \emph{Handbook of
  {{Applied Spatial Analysis}}: {{Software Tools}}, {{Methods}} and
  {{Applications}}}, pages 175--193. {Springer}, {Berlin, Heidelberg}, 2010.
\newblock ISBN 978-3-642-03647-7.
\newblock \doi{10.1007/978-3-642-03647-7_11}.

\bibitem[Lesage and Fischer(2008)]{lesageSpatialGrowthRegressions2008}
James~P. Lesage and Manfred~M. Fischer.
\newblock Spatial {{Growth Regressions}}: {{Model Specification}},
  {{Estimation}} and {{Interpretation}}.
\newblock \emph{Spatial Economic Analysis}, 3\penalty0 (3):\penalty0 275--304,
  November 2008.
\newblock ISSN 1742-1772.
\newblock \doi{10.1080/17421770802353758}.

\bibitem[Anselin and Rey(2014)]{anselinModernSpatialEconometrics2014}
Luc Anselin and Sergio~J. Rey.
\newblock \emph{Modern Spatial Econometrics in Practice: A Guide to {{GeoDa}},
  {{GeoDaSpace}} and {{PySAL}}}.
\newblock {GeoDa Press LLC, Chicago, IL}, 2014.
\newblock ISBN 978-0-9863421-0-3.

\bibitem[{GuoMeng} et~al.(2021){GuoMeng}, {cheng}, and
  Shukai]{guomengAnalysisDirectEffect2021}
{GuoMeng}, Du~Zhong {cheng}, and Cai Shukai.
\newblock An analysis of the ``direct effect'' and ``indirect effect'' of urban
  housing prices on the upgrading of industrial
  structure\textemdash\textemdash{{Based}} on data of 285 cities.
\newblock \emph{Journal of Physics: Conference Series}, 1792\penalty0
  (1):\penalty0 012017, February 2021.
\newblock ISSN 1742-6588, 1742-6596.
\newblock \doi{10.1088/1742-6596/1792/1/012017}.

\bibitem[Bivand and Piras(2015)]{bivandComparingImplementationsEstimation2015}
Roger Bivand and Gianfranco Piras.
\newblock Comparing {{Implementations}} of {{Estimation Methods}} for {{Spatial
  Econometrics}}.
\newblock \emph{Journal of statistical software}, 63:\penalty0 1--36, February
  2015.
\newblock \doi{10.18637/jss.v063.i18}.

\bibitem[Kendall(1938)]{kendallNewMeasureRank1938}
M.~G. Kendall.
\newblock A new measure of rank correlation.
\newblock \emph{Biometrika}, 30\penalty0 (1):\penalty0 81--93, 1938.
\newblock ISSN 0006-3444.
\newblock \doi{10.2307/2332226}.
\newblock URL \url{https://www.jstor.org/stable/2332226}.
\newblock Publisher: [Oxford University Press, Biometrika Trust].

\bibitem[LeSage and Pace(2009)]{lesageIntroductionSpatialEconometrics2009a}
James LeSage and Robert~Kelley Pace.
\newblock \emph{Introduction to {{Spatial Econometrics}}}.
\newblock {Chapman and Hall/CRC}, {New York}, January 2009.
\newblock ISBN 978-0-429-13808-9.
\newblock \doi{10.1201/9781420064254}.

\bibitem[Anselin(1988)]{anselinSpatialEconometricsMethods1988}
Luc Anselin.
\newblock \emph{Spatial {{Econometrics}}: {{Methods}} and {{Models}}}, volume~4
  of \emph{Studies in {{Operational Regional Science}}}.
\newblock {Springer Netherlands}, {Dordrecht}, 1988.
\newblock ISBN 978-90-481-8311-1 978-94-015-7799-1.
\newblock \doi{10.1007/978-94-015-7799-1}.

\bibitem[De~Palma and Papageorgiou(1988)]{depalmaHeterogeneityStatesUrban1988}
Andr{\'e} De~Palma and Yorgos~Y. Papageorgiou.
\newblock Heterogeneity in states and urban structure.
\newblock \emph{Regional Science and Urban Economics}, 18\penalty0
  (1):\penalty0 37--56, February 1988.
\newblock ISSN 01660462.
\newblock \doi{10.1016/0166-0462(88)90004-X}.

\bibitem[Jia et~al.(2019)Jia, Du, Wang, Bai, and
  Fei]{jiaMeasuringVibrancyUrban2019}
Chen Jia, Yunyan Du, Siying Wang, Tianyang Bai, and Teng Fei.
\newblock Measuring the vibrancy of urban neighborhoods using mobile phone data
  with an improved {{PageRank}} algorithm.
\newblock \emph{Transactions in GIS}, 23\penalty0 (2):\penalty0 241--258, 2019.
\newblock ISSN 1467-9671.
\newblock \doi{10.1111/tgis.12515}.

\end{thebibliography}
